\documentclass[aps,pre,10pt,groupedaddress,amsfonts,amssymb,amsmath,showpacs,showkeys,a4paper,floatfix,notitlepage]{revtex4-1}

\usepackage[english]{babel}
\usepackage[T1]{fontenc}
\usepackage{lmodern}
\usepackage[utf8]{inputenc}
\usepackage[protrusion=true,expansion=true]{microtype}
\usepackage{amsmath,amsfonts,amsthm}
\usepackage{ bbold }
\usepackage[pdftex]{graphicx}
\usepackage[outdir=./]{epstopdf}
\usepackage{bm}
\usepackage[usenames,dvipsnames,svgnames,table]{xcolor}
\usepackage{csquotes}
\usepackage{booktabs} 
\usepackage{array} 
\usepackage{geometry}
\usepackage{MnSymbol}
\usepackage{pgf}
\usepackage{import}
\usepackage{caption}
\usepackage{subcaption}
\usepackage{wrapfig} 
\usepackage[normalem]{ulem} 
\usepackage{comment} 

\newif\ifchanges
\changestrue

\definecolor{lbgk_color}{rgb}{0,0,0}
\definecolor{kbcA_color}{HTML}{377EB8}
\definecolor{kbcB_color}{HTML}{4DAF4A}
\definecolor{kbcC_color}{HTML}{984EA3}
\definecolor{kbcD_color}{HTML}{FF7F00}
\definecolor{entropic_color}{HTML}{E41A1C}

\newcommand{\lbgkclr}[1]{{\color{lbgk_color}#1}}
\newcommand{\kbcAclr}[1]{{\color{kbcA_color}#1}}
\newcommand{\kbcBclr}[1]{{\color{kbcB_color}#1}}
\newcommand{\kbcCclr}[1]{{\color{kbcC_color}#1}}
\newcommand{\kbcDclr}[1]{{\color{kbcD_color}#1}}
\newcommand{\elbmclr}[1]{{\color{entropic_color}#1}}

\newcommand{\lbgksym}{\lbgkclr{solid}}
\newcommand{\kbcAsym}{\kbcAclr{dashed}}
\newcommand{\kbcBsym}{\kbcBclr{$\bm\bigtimes$}}
\newcommand{\kbcCsym}{\kbcCclr{dotted}}
\newcommand{\kbcDsym}{\kbcDclr{$\bm\medsquare$}}
\newcommand{\elbmsym}{\elbmclr{$\bm\medcircle$}}

\newcommand{\feq}{f^{\rm eq}}
\newcommand{\fmirr}{f^{\rm mirr}}

\newcommand{\integral}[1]{\left\langle{#1}\right\rangle}

\begin{document}

\title{Entropic Multi-Relaxation Models for Simulation of Fluid Turbulence}

\thanks{To be published in Proceedings of Discrete Simulation of Fluid Dynamics DSFD 2014.}

\author{Fabian B\"osch}
\email{boesch@lav.mavt.ethz.ch}
\affiliation{Department of Mechanical and Process Engineering, ETH Zurich, 8092 Zurich, Switzerland}

\author{Shyam S. Chikatamarla}
\affiliation{Department of Mechanical and Process Engineering, ETH Zurich, 8092 Zurich, Switzerland}

\author{Ilya Karlin}
\affiliation{Department of Mechanical and Process Engineering, ETH Zurich, 8092 Zurich, Switzerland}

\date{\today}

\begin{abstract} A recently introduced family of lattice Boltzmann (LB) models (Karlin, B\"osch, Chikatamarla, Phys. Rev. E, 2014; Ref\cite{KBC})
is studied in detail for incompressible two-dimensional flows. A framework for developing LB models based on entropy considerations is laid out extensively. Second order rate of convergence is numerically confirmed and it is demonstrated that these entropy based models recover the Navier-Stokes solution in the hydrodynamic limit. Comparison with the standard Bhatnagar-Gross-Krook (LBGK) and the entropic lattice Boltzmann method (ELBM) demonstrates the superior stability and accuracy for several benchmark flows and a range of grid resolutions and Reynolds numbers. High Reynolds number regimes are investigated through the simulation of two-dimensional turbulence, particularly for under-resolved cases. 
Compared to resolved LBGK simulations, the presented class of LB models demonstrate excellent performance and capture the 
turbulence statistics with good accuracy.\end{abstract}

\maketitle

\section{Introduction}

In recent years, the lattice Boltzmann (LB) method has made substantial progress towards a
successful and particularly efficient approach to computational fluid dynamics. 
By employing a system of discrete kinetic equations rather than solving directly the macroscopic flow equations it has shown its potential in a wide 
range of applications, from turbulence phenomena \cite{benzi1990two,TURBO} to flows at a micron scale \cite{Ansumali2007} and multiphase flows  \cite{Swift1995}, to relativistic hydrodynamics \cite{Miller1}, soft-glassy systems \cite{Benzi2009} and beyond.

The kinetic system describes the discrete-time dynamics of populations $f_i(\bm{x},t)$ which are designed to reproduce the Navier-Stokes equations in the hydrodynamic limit.
Each population $f_i$ corresponds to a discrete microscopic velocity $\bm{v}_i$, $i=1,\dots,b$, which fits into a regular spatial lattice with the nodes $\bm{x}$. This enables a highly efficient `stream-along-links-and-equilibrate-at-nodes' realization of the LB algorithm. 
We consider the {single-phase} isothermal case in two dimensions for the purpose or this paper.
A general form of the LB equation can be written as
\begin{equation}
\label{eq:Egeneral}
f_i(\bm{x}+\bm{v}_i,t+1)=f_i'\equiv(1-\beta)f_i(\bm{x},t)+\beta f_i^{\rm mirr}(\bm{x},t).
\end{equation}
Here the left-hand side is the propagation of the populations along the lattice links, while the right-hand side is the so-called {post-collision} state $f'$. The {post-collision} state is a convex linear combination between the {pre-collision} state $f$ and the mirror state $\fmirr$. The choice of $\fmirr$ as
\begin{equation}
\label{eq:LBGKmirr}
f_i^{\rm mirr}=2f_i^{\rm eq}-f_i
\end{equation}
results in the well known LBGK model
which has most notably led to the success of the method, in particular for the simulation of incompressible flows \cite{SucciRev,Chen92,Qian92,ChenDoolen} and complex hydrodynamic phenomena \cite{succi,Aidun10}.

The local equilibrium $f_i^{\rm eq}$ is found as a maximizer of the entropy,
\begin{equation}
\label{eq:S}
S[f]=-\sum_{i=1}^b f_i\ln\left(\frac{f_i}{W_i}\right),
\end{equation}
subject to fixed locally conserved fields, $\rho=\sum_{i=1}^b f_i$ (density) and $\rho\bm{u}=\sum_{i=1}^b \bm{v}_if_i$ (momentum density), and where the weights $W_i$ are lattice-specific constants. 
The equilibrium can be approximated in closed form to second order in velocity as
\begin{equation}
\label{eq:Eq}
\feq_i = W_i \rho \left[1 + \tfrac{u_\alpha v_{i\alpha}}{c_s^2}  + \tfrac{u_\alpha u_\beta(v_{i\alpha} v_{i\beta} - c_s^2 \delta_{\alpha\beta})}{2 c_s^4}  \right] + O(u^3)
\end{equation}
where $c_{\rm s}$ is the speed of sound (a lattice dependent $O(1)$ constant).
The LBGK equations, \eqref{eq:Egeneral} and (\ref{eq:LBGKmirr}), recover the Navier-Stokes equation for the fluid velocity $\bm{u}$ in the hydrodynamic limit, under the assumption $u \ll c_s$, while the relaxation parameter $\beta \in [0,1]$ is defined by the kinematic viscosity $\nu$ 
\begin{equation}
\label{eq:viscosity}
\nu=c_s^2\left(\frac{1}{2\beta}-\frac{1}{2}\right).
\end{equation}
In order to achieve high Reynolds numbers, one must decrease viscosity as the flow velocity is restricted to low Mach numbers by construction of the kinetic system and, thus, the limit $\beta\to 1$ is of great importance. 
While the time step and the lattice spacing are connected by the relation $\delta x_i = \delta t v_i$ (where $\delta x_i$ is the lattice spacing in the direction of discrete velocity $v_i$), the kinematic viscosity can be chosen independently of the spatial and temporal resolution. This feature makes the LBGK model particularly attractive for high Reynolds number simulations, if only in principle.

Despite of its promising nature and popularity, however, the LBGK model shows numerical instabilities already at moderate Reynolds numbers unless a rather high resolution is employed, which quickly becomes computationally prohibitive. This precluded the LB method from making a sustainable impact in the field of computational fluid dynamics.

A number of approaches can be found in the literature intended to alleviate this issue. We will restrict the following short discussion to methods without explicit turbulence models. Most notably, the entropic lattice Boltzmann method (ELBM) features non-linear stability and has shown excellent performance \cite{Karlin99,cylinder,CK_PRL1}. While ELBM converges to LBGK in the resolved case, it locally modifies the relaxation rate which in turn modifies the viscosity in order to fulfil the second law of thermodynamics by both enhancing and smoothing the features of the flow where necessary subject to an entropy condition. Another approach using multiple relaxation parameters (MRT) is widely used, which does not affect viscosity in the first place but requires careful tuning of the relaxation parameters. Although MRT models were successful in slightly stabilizing the LB method, they still remain challenged by high Reynolds numbers \cite{Freitas2011}.

Recently, the authors have developed a scheme without the need for tunable parameters or turbulent viscosity (Karlin, B\"osch, Chikatamarla, Phys. Rev. E 2014; Ref\cite{KBC})
which has demonstrated a significant extension in the operation range for simulations at high Reynolds numbers. Promising results have been reported for both two and three dimensions, as well as for complex boundaries and in presence of turbulence. Much alike ELBM, entropic considerations have been employed to render the scheme stable without introducing considerable computational overhead and by keeping the simplicity and locality of the LBGK and MRT schemes. Below we shall refer to this class of models as KBC models 
for brevity.

While in \cite{KBC} one particular realization of the model was discussed, this paper focuses on four variations of the KBC family which will be investigated in detail, both numerically and analytically. 
We restrict ourselves to two dimensional fully periodic domains in the absence of wall boundaries in order to assess the stability and accuracy of the scheme independently of the errors arising from the wall.

The MRT class of LB models separate the relaxation into various groups based on separation of scales between the fast and slow varying moments. Moreover, since the relaxation of the {off-diagonal} parts of the pressure tensor are fixed by the choice of kinematic viscosity, the MRT scheme asserts that the relaxation of higher order moments should not affect the flow field (up to the Navier-Stokes level) and hence can be used to construct more stable LB schemes. Following this line of thinking, several MRT schemes  were suggested for the choice of relaxation of higher order moments (beyond the pressure tensor) \cite{HSB,dHumieres92,Geier2006}. The KBC models extend this idea using local entropy considerations and demonstrate that much higher Reynolds numbers can be achieved on much smaller grid sizes. 

Three well-studied benchmark flows - Green-Taylor vortex, doubly periodic shear layer and decaying two-dimensional turbulence - are simulated for all the four KBC models as well as for the classical LBGK and the non-linearly stable ELBM. Results are compared to each other as well as to reference solutions in order to evaluate the models.

\section{Moment Representation}

We consider the standard nine-velocity model (D2Q9). The discrete velocities are constructed as a tensor product of two one-dimensional velocity sets, $v_{(i)}=i$, where $i=0,\pm 1$; thus $v_{(i,j)}=(v_{(i)},v_{(j)})$ in the fixed Cartesian reference frame. 

We recall that any product lattice, such as the D2Q9, is characterized  by  natural moments. For D2Q9, these natural moments are $\rho M_{pq}$, where $\rho=\integral{f_{(i,j)}}$ is the density, and
\begin{equation}
\label{eq:natmoments}
\rho M_{pq}=\langle f_{(i,j)}v_{(i)}^p v_{(j)}^q\rangle,\ p,q\in\{0,1,2\}.
\end{equation}
In the sequel we use the following linear combinations to represent natural moments (\ref{eq:natmoments})
\begin{equation}\label{eq:natmoments1}
\begin{split}
M_{00}, u_x=M_{10},\ u_y=M_{01},\ T=M_{20}+M_{02},\ N=M_{20}-M_{02},\ \Pi_{xy}=M_{11},\\
Q_{xyy}=M_{12},\ Q_{yxx}=M_{21},\ A=M_{22}.
\end{split}
\end{equation}
These are interpreted as the normalization to the density ($M_{00}=1$), the flow velocity components ($u_x$, $u_y$), the trace of the pressure tensor at unit density ($T$), the normal stress difference at unit density ($N$), and the  off-diagonal component of the pressure tensor at unit density ($\Pi_{xy}$). The (linearly independent) third-order moments ($Q_{xyy}$, $Q_{yxx}$) and the fourth-order moment ($A$) lack a direct physical interpretation for incompressible flows.

The $b$ linearly independent moments serve as a different basis for the kinetic equations. It is clear that the choice of basis vectors is not unique. Another basis is given by the central moments of the form
\begin{equation}
\rho\tilde{M}_{pq}=\integral{(v_{(i)}-u_x)^p(v_{(j)}-u_y)^q f_{(i,j)}},
\end{equation}
where we have the following relation between the natural and the central moments
\begin{align}\label{eq:cm}
\begin{split}
&\Pi_{xy}=\tilde{\Pi}_{xy}+u_xu_y,\\
&N=\tilde{N}+(u_x^2-u_y^2),\\
&T=\tilde{T}+u^2,\\
&Q_{xyy}=\tilde{Q}_{xyy}+2u_y\tilde{\Pi}_{xy}-\tfrac{1}{2}u_x\tilde{N}+\tfrac{1}{2}u_x\tilde{T}+u_x u_y^2,\\
&Q_{yxx}=\tilde{Q}_{yxx}+2u_x\tilde{\Pi}_{xy}+\tfrac{1}{2}u_y\tilde{N}+\tfrac{1}{2}u_y\tilde{T}+u_y u_x^2,\\
&A=\tilde{A} +2\left[u_x\tilde{Q}_{xyy}+u_y\tilde{Q}_{yxx}\right]
+4u_xu_y\tilde{\Pi}_{xy}+\tfrac{1}{2}u^2 \tilde{T}-\tfrac{1}{2}(u_x^2-u_y^2)\tilde{N}+u_x^2u_y^2.\\
\end{split}
\end{align}
We remark in passing that the mapping of natural moments onto central moments is nonlinear (it explicitly depends on the powers of the velocity components).

It is important to note that the macroscopic equations are recovered by a projection of the kinetic system onto the lower order moments $\rho, u_x, u_y$. The higher order moments are thus in our hands, in principle. However, they play an important role for the numerical stability of the scheme. Precisely this observation has been widely used to construct models with the goal to stabilize the LB scheme, among which are the MRT and KBC models.

With the set of natural moments (\ref{eq:natmoments1}), populations are uniquely represented as follows ($\sigma,\lambda=\{-1,1\}$):
\begin{align}
\begin{split}\label{eq:momentsD2Q9}
f_{(0,0)}&=\rho\left(1-T+A\right),\\
f_{(\sigma,0)}&=\tfrac{1}{2}\rho\left(\tfrac{1}{2}(T+N)+\sigma u_x-\sigma Q_{xyy}-A\right),\\
{f}_{(0,\lambda)}&=\tfrac{1}{2}\rho\left(\tfrac{1}{2}(T-N)+\lambda u_y-\lambda Q_{yxx}-A\right),\\
f_{(\sigma,\lambda)}&=\tfrac{1}{4}\rho\left(A+(\sigma)(\lambda)\Pi_{xy}+\sigma Q_{xyy}+\lambda Q_{yxx}\right).
\end{split}
\end{align}
A similar representation can be written using the central moments by substituting eq. \eqref{eq:cm} in \eqref{eq:momentsD2Q9}.

We group the population's natural and central moment representations into the following functions for convenience: $k_i$, $t_i(T)$, $n_i(N)$, $p_i(\Pi_{xy})$, $q_i(Q_{xyy},Q_{yxx})$, $a_i(A)$ and $\tilde{k}_i$, $\tilde{t}_i(\tilde{T})$, $\tilde{n}_i(\tilde{N})$, $\tilde{p}_i(\tilde{\Pi}_{xy})$, $\tilde{q}_i(\tilde{Q}_{xyy},\tilde{Q}_{yxx})$, $\tilde{a}_i(\tilde{A})$, respectively, where each of these groups also depends on the density and velocity, $\rho, \bm{u}$, of the flow. $k_i(\rho,\bm{u})$ and $\tilde{k}_i(\rho,\bm{u})$ represent the kinematic part only.
Thus, the populations are rewritten as
\begin{equation}
f_i = k_i + t_i(T) + n_i(N) + p_i(\Pi_{xy}) + q_i(Q_{xyy},Q_{yxx}) + a_i(A);
\end{equation}
shown here in the natural moment representation.
With this formulation, the mirror state \eqref{eq:LBGKmirr} can be redefined by introducing relaxation parameters $\gamma_k$
\begin{equation}
\label{eq:general_mirror}
\begin{split}
f_i^{\rm mirr} = k_i + 
\left[\gamma_0 t_i(T^{\rm eq}) + (1-\gamma_0)t_i(T) \right] + 
\left[\gamma_1 n_i(N^{\rm eq}) + (1-\gamma_1)n_i(N) \right] \\ + 
\left[\gamma_2 p_i(\Pi_{xy}^{\rm eq}) + (1-\gamma_2)p_i(\Pi_{xy}) \right] + 
\left[\gamma_3 q_i(Q_{xyy}^{\rm eq},Q_{yxx}^{\rm eq}) + (1-\gamma_3)q_i(Q_{xyy},Q_{yxx}) \right] \\+ 
\left[\gamma_4 a_i(A^{\rm eq}) + (1-\gamma_4)a_i(A) \right].
\end{split}
\end{equation}
We now need to find the optimal values for the relaxation parameters $\gamma_k$. Let us first note that the relaxation parameters for the parts depending on the entries of the stress tensor, $n_i$, $p_i$, must be set to $\gamma_1=\gamma_2=2$ in order to reproduce the Navier-Stokes equations, at least to second order.
The relaxation for remaining moments can be chosen without affecting the hydrodynamic limit. 

A seemingly natural choice would be to set these relaxation parameters to $\gamma_{i} = 1/\beta$ (that is, $\gamma_i \approx 1$ for $\beta \rightarrow 1$) for $i \in \lbrace 0,3,4 \rbrace$ which implies that all higher order moments are brought to their equilibrium. This is essentially the idea of regularized lattice Boltzmann model \cite{Latt,Chen2006125,Montessori2014}. However, in many benchmark simulations this choice does not lead to significantly better results as compared to the LBGK.
MRT models on the other hand suggest highly optimized but fixed values for the relaxation of higher order moments. 

The KBC models take a different route by making the relaxation of higher order moments adapt to the flow and by letting local entropy decide about the corresponding relaxation values. 
We will discuss here KBC models with only one free relaxation parameter where the populations are represented as sum of three moment functions
\begin{equation}
\label{eq:ksh-representation}
f_i=k_i+s_i+h_i,
\end{equation}
where $k_i$ (= kinematic part) depends only on the locally conserved fields, $s_i$ (= shear part) depends on the stress tensor 
$\bm{\Pi}=\sum_{i=1}^b\bm{v}_i\otimes\bm{v}_if_i$, 
and $h_i$ (= higher-order moments) is a linear combination of the remaining higher-order moments. In the presentation \eqref{eq:ksh-representation} we essentially lump together all the higher order moments. Further extensions can be envisaged by splitting the higher order moments, $h_i$ into individual components in order to further improve the stability. However, we show in this paper that even with the suggested lumping of moments $h_i$ extremely stable LB models can be readily created which outperform MRT models. 

With the representation (\ref{eq:ksh-representation}), the mirror state is in a one-parameter form,
\begin{equation}\label{eq:ksh-mirror}
f_i^{\rm mirr}=k_i+[2s_i^{\rm eq}-s_i]+[\gamma h_i^{\rm eq}+(1-\gamma)h_i],
\end{equation}
where $\gamma$ is a relaxation parameter which is not yet specified. For $\gamma=2$, KBC \eqref{eq:ksh-mirror} coincides with LBGK. 
For any $\gamma$, the resulting LB model still recovers hydrodynamics with the same kinematic viscosity 
$\nu$ (\ref{eq:viscosity}) which is demonstrated in section \ref{sec:hydro}. 

\begin{table}
\centering
\begin{tabular}{lll}
\toprule
Model    & $s(\,\cdot\,;\rho,\bm{u})$ & $h(\,\cdot\,;\rho,\bm{u})$ \\
\midrule
LBGK      & $\Pi_{xy}, N, T, Q_{xyy}, Q_{yxx}, A$     & -      \\
KBC A     & $\tilde{\Pi}_{xy},\tilde{N}, \tilde{T}$    & $\tilde{Q}_{xyy},\tilde{Q}_{yxx},\tilde{A}$      \\
KBC B     & $\tilde{\Pi}_{xy},\tilde{N}$               & $\tilde{T}, \tilde{Q}_{xyy},\tilde{Q}_{yxx},\tilde{A}$      \\
KBC C     & $\Pi_{xy},N, T$                            & $Q_{xyy},Q_{yxx},A$      \\
KBC D     & $\Pi_{xy},N$                               & $T, Q_{xyy},Q_{yxx},A$      \\
\bottomrule
\end{tabular}
\caption{Moment grouping for discussed models.}
\label{tab:kbc_moment_grouping}
\end{table}

\section{Model Description}

Let us define the four KBC models A, B, C and D which we consider here. Models A and B are represented in the basis spanned by central moments while C and D are represented by natural moments. Models B and D, however, differ from A and C by including the moment function depending on the trace of the stress tensor, $\tilde{t}_i(\tilde{T})$ and $t_i(T)$, respectively, in the higher order part $h$. This eventually leads to different coefficients for the bulk viscosity. Table \ref{tab:kbc_moment_grouping} summarizes the contribution for $s$ and $h$ for the different models.

\section{Entropic Stabilizer}

We will review the definition of the entropic stabilizer $\gamma$ given by \cite{KBC} in the following.
Let $S(\gamma)$ be the entropy of the post-collision state appearing on the right hand side of (\ref{eq:Egeneral}), with the mirror state (\ref{eq:ksh-mirror}). We require the stabilizer $\gamma$ to correspond to the maximum of this function.
Introducing deviations $\Delta s_i=s_i-s_i^{\rm eq}$ and $\Delta h_i=h_i-h_i^{\rm eq}$, the condition for the critical point reads:
\begin{equation}\label{eq:result1}
\sum_{i=1}^{b}\Delta h_i\ln\!\left(1+\frac{(1-\beta\gamma)\Delta h_i-(2\beta-1)\Delta s_i}{f_i^{\rm eq}}\right)=0.
\end{equation}
Equation (\ref{eq:result1}) suggests that among all non-equilibrium states with the fixed mirror values of the stress, 
$s_i^{\rm mirr}=2s_i^{\rm eq}-s_i$, we pick the one which maximizes the entropy. 
Note that $\gamma$ self-adapts to a value given by the maximum entropy condition at each grid point (\ref{eq:result1}) and thus eliminates the need for tuning. We observe in all our simulation that the system entropy is monotonically growing over time. It is therefore conjectured that the second law of thermodynamics is fulfilled in practice and moreover, by providing a Lyapunov function, the entropy based relaxation contributes crucially to the stability of the scheme.

An estimate for $\gamma$ in a closed form can be obtained by introducing the  entropic scalar product $\langle X{|}Y\rangle$ in the $b$-dimensional vector space,
\begin{equation}
\langle X{|}Y\rangle=\sum_{i=1}^{b}\frac{X_iY_i}{f_i^{\rm eq}},
\end{equation}
and expanding (\ref{eq:result1}) to the first non-vanishing order in $\Delta s_i/f_i^{\rm eq}$ and $\Delta h_i/f_i^{\rm eq}$ which yields
\begin{equation}
\label{eq:result21}
\gamma^*= \frac{1}{\beta}-\left(2-\frac{1}{\beta}\right)\frac{\langle\Delta s{|}\Delta h\rangle}{\langle\Delta h{|}\Delta h\rangle}
\end{equation}
This estimate has proven to be sufficient for all practical purposes by stabilizing the scheme and monotonically incrementing the system entropy $S$. It is used for all the simulations presented in this work.

Unlike any MRT, the relaxation parameter $\gamma$ is neither fixed a priori nor is it constant in space and time for KBC models. Therefore, it is instructive to obtain an estimate on the asymptotics for the different models. Let us expand $\langle\Delta s{|}\Delta h\rangle$ in powers of velocity $\bm{u}$, $\langle\Delta s{|}\Delta h\rangle = \langle\Delta s{|}\Delta h\rangle^{(0)} + \langle\Delta s{|}\Delta h\rangle^{(1)} + \cdots$. At zeroth order this leads to
\begin{align}
\langle \Delta s_A{|}\Delta h_A\rangle^{(0)} &= -\tfrac{1}{4} \rho (9 \tilde{A}-1) (3 \tilde{T}-2)\\
\label{eq:dsdh0B}
\langle \Delta s_B{|}\Delta h_B\rangle^{(0)} &= 0 \\
\langle \Delta s_C{|}\Delta h_C\rangle^{(0)} &= -\tfrac{1}{4} \rho (9 \tilde{A}-1) (3 \tilde{T}-2)\\
\langle \Delta s_D{|}\Delta h_D\rangle^{(0)} &=0
\end{align}
where we have chosen to replace natural moments by central moments according to \eqref{eq:cm}. Note that the two models, B and D, which include the trace of the stress tensor in the higher-order part $h$ do not contribute to the leading order. The next order contributions are 

\begin{align}
\langle \Delta s_A{|}\Delta h_A\rangle^{(1)} &= 0 \\
\label{eq:dsdh1B}
\langle \Delta s_B{|}\Delta h_B\rangle^{(1)} &= 0 \\
\langle \Delta s_C{|}\Delta h_C\rangle^{(1)} &= \tfrac{9}{4}\rho (\tilde{Q}_{xyy} ((2+3 \tilde{N}-3 \tilde{T}) u_x-12 \tilde{P}_{xy} u_y)-\tilde{Q}_{yxx} (12 \tilde{P}_{xy} u_x+(-2+3 \tilde{N}+3 \tilde{T}) u_y)) \\
\langle \Delta s_D{|}\Delta h_D\rangle^{(1)} &=-\tfrac{27}{4} \rho (-\tilde{N} \tilde{Q}_{xyy} u_x+4 \tilde{P}_{xy} \tilde{Q}_{yxx} u_x+4 \tilde{P}_{xy} \tilde{Q}_{xyy} u_y+\tilde{N} \tilde{Q}_{yxx} u_y).
\end{align}
We see that KBC B is the only model among the four that does not contribute to either zeroth or first order. This leads to the conjecture that, when $\beta \rightarrow 1$, $\gamma$ will be close to $1$ for the KBC B model, unlike the other KBC models. 
This is clearly confirmed by our simulation results (see, e.g. figs. \ref{fig:2dturbPalinGammaR13} and \ref{fig:2dturbEnergyR150}). 
The average value of the entropic stabilizer $\gamma \approx 1$ is found in our simulations for a range of Reynolds numbers and resolutions. Note that by fixing $\gamma=1/\beta$ the KBC models coincide with the regularized LB model (more specifically, with a realization of the regularized LB in the according moment basis). Thus, the above theoretical derivation of the entropic stabilizer for KBC model B produced the empirical regularized LB in the central moment basis.

\section{Hydrodynamic Limit of KBC models}
\label{sec:hydro}

Let us derive the hydrodynamic limit of the general kinetic equation with KBC-type mirror state $\fmirr$ \eqref{eq:ksh-mirror}. 
We start by rewriting eq.\ (\ref{eq:Egeneral}) and eq.\ (\ref{eq:ksh-mirror}) as 
\begin{equation}
\label{eq:EGE}
f_i' = f_i + 2\beta\left(f^{\rm GE}_i - f_i\right).
\end{equation}
with the generalized equilibrium  \cite{Asinari09,Asinari10,Asinari11} of the form,
\begin{equation}
\label{eq:GE}
f_i^{\rm GE} = \feq_i + \tfrac{1}{2}(\gamma-2)(h^{\rm eq}_i - h_i).
\end{equation}

In the following derivation, Einstein's summation convention is applied for all subscript indices except for $i$ where the explicit notation $\integral{...}$ is used.
It is useful to compute the second and third order equilibrium moment until second order in velocity beforehand
\begin{align}
\label{eq:Peq}
\Pi_{\alpha\beta}^{\rm eq} &\equiv \integral{\feq_i v_{i\alpha}v_{i\beta}} = \rho c_s^2 \delta_{\alpha\beta} + \rho u_\alpha u_\beta, \\
\label{eq:Qeq}
Q_{\alpha\beta\mu}^{\rm eq} &\equiv \integral{\feq_i v_{i\alpha}v_{i\beta}v_{i\mu}} = \rho c_s^2\left( u_\alpha\delta_{\beta\mu} + u_\beta\delta_{\alpha\mu} +  u_\mu\delta_{\alpha\beta} \right).
\end{align}
Due to local conservation laws and as a direct consequence of the construction of the moment groups $k$, $s$ and $h$ (see eq. \eqref{eq:ksh-representation} and table \ref{tab:kbc_moment_grouping}) we can immediately state the following relations for the zeroth and first order moments
\begin{align}
\label{eq:k_conservation}
\integral{k_i \lbrace 1, v_{i\alpha} \rbrace } &= \lbrace \rho, \rho u_\alpha \rbrace, \\
\label{eq:sh_conservation}
\integral{s_i \lbrace 1, v_{i\alpha} \rbrace } = \integral{h_i \lbrace 1, v_{i\alpha} \rbrace } &= 0.
\end{align}
While these relations hold for all four KBC models discussed here, they depart from each other in the higher order moments. Table \ref{tab:sh_moments2} shows the second order moments for moment functions $s$ and $h$, respectively. Note that all the higher order moments for the kinematic part $k$ vanish.

\begin{table}
\centering
\begin{tabular}{lcccccc}
\toprule
Model    & 
$\integral{s_i v_{ix} v_{ix}}$ & $\integral{s_i v_{iy} v_{iy}}$ & $\integral{s_i v_{ix} v_{iy}}$ & 
$\integral{h_i v_{ix} v_{ix}}$ & $\integral{h_i v_{iy} v_{iy}}$ & $\integral{h_i v_{ix} v_{iy}}$ \\
\midrule
KBC A     & 
$\tfrac{1}{2}\rho(\tilde{T}+\tilde{N})$ & $\tfrac{1}{2}\rho(\tilde{T}-\tilde{N})$ & $\rho\tilde{\Pi}_{xy}$ &
$0$                                     & $0$                                     & $0$ \\
KBC B     & 
$\tfrac{1}{2}\rho\tilde{N}$             & $-\tfrac{1}{2}\rho\tilde{N}$            & $\rho\tilde{\Pi}_{xy}$ &
$\tfrac{1}{2}\rho\tilde{T}$             & $\tfrac{1}{2}\rho\tilde{T}$             & $0$ \\
KBC C     & 
$\tfrac{1}{2}\rho(T+N)$                 & $\tfrac{1}{2}\rho(T-N)$                 & $\rho\Pi_{xy}$ &
$0$                                     & $0$                                     & $0$ \\
KBC D     & 
$\tfrac{1}{2}\rho N$    & $-\tfrac{1}{2}\rho N$ & $\rho \Pi_{xy}$ &
$\tfrac{1}{2}\rho T$    & $\tfrac{1}{2}\rho T$    & $0$ \\
\bottomrule
\end{tabular}
\caption{Second order moments for functions $s$ and $h$ in KBC.}
\label{tab:sh_moments2}
\end{table}

After the previous preliminary considerations let us expand the left hand side of equation \eqref{eq:EGE} using a Taylor series to second order
\begin{equation}
\label{eq:taylor}
\left[ \delta t(\partial_t + \partial_\alpha v_{i\alpha}) + \tfrac{\delta t^2}{2}(\partial_t + \partial_\alpha v_{i\alpha})(\partial_t + \partial_\beta v_{i\beta}) \right] f_i = 2\beta\left[ \feq_i - f_i + \tfrac{1}{2}(\gamma-2)(h^{\rm eq}_i - h_i) \right].
\end{equation}
By introducing a characteristic time scale of the flow, $\Theta$, we can rewrite \eqref{eq:taylor} in a non-dimensional form using reduced variables $t'=t/\Theta$, $v_i' = v_i/c$ and $x' = x/(c\Theta)$, where $c=1$. After introduction of the parameter $\epsilon = \delta t/\Theta$ and dropping the primes to simplify notation we get
\begin{equation}
\label{eq:taylor_nondim}
\left[ \epsilon(\partial_t + \partial_\alpha v_{i\alpha}) + \tfrac{\epsilon^2}{2}(\partial_t + \partial_\alpha v_{i\alpha})(\partial_t + \partial_\beta v_{i\beta}) \right] f_i = 2\beta\left[ \feq_i - f_i + \tfrac{1}{2}(\gamma-2)(h^{\rm eq}_i - h_i) \right].
\end{equation}

By exploiting the smallness of $\epsilon$ we can perform a multiscale expansion of the time derivative operator, the populations and their decomposition into $s$ and $h$ parts until second order,
\begin{align}
\label{eq:expand_dt}
\epsilon \partial_t &= \epsilon \partial_t^{(1)} + \epsilon^2\partial_t^{(2)} + \cdots \\
\label{eq:expand_f}
f_i &= f_i^{(0)} + \epsilon f_i^{(1)} + \epsilon^2 f_i^{(2)}  + \cdots \\
\label{eq:expand_s}
s_i &= s_i^{(0)} + \epsilon s_i^{(1)} + \epsilon^2 s_i^{(2)}  + \cdots \\
\label{eq:expand_h}
h_i &= h_i^{(0)} + \epsilon h_i^{(1)} + \epsilon^2 h_i^{(2)}  + \cdots.
\end{align}

Inserting~eqs.~(\ref{eq:expand_dt}) to~(\ref{eq:expand_h})
into eq.~\eqref{eq:taylor_nondim} we can analyze the terms corresponding to orders $\epsilon^0$, $\epsilon^1$ and $\epsilon^2$. The zeroth order terms lead to
\begin{equation}
\label{eq:zeroth_order}
0 = 2\beta\left[ \feq_i - f_i^{(0)} + \tfrac{1}{2}(\gamma-2)(h^{\rm eq}_i - h_i^{(0)}) \right],
\end{equation}
which implies
\begin{equation}
\label{eq:f_h_zeroth_order}
f_i^{(0)} = \feq_i,~~h_i^{(0)} = h_i^{\rm eq}.
\end{equation}
Local conservation laws dictate the relations $\integral{f_i \lbrace 1, v_{i\alpha} \rbrace} = \integral{\feq_i \lbrace 1, v_{i\alpha} \rbrace}$ and\\ $\integral{h_i \lbrace 1, v_{i\alpha} \rbrace} = \integral{h^{\rm eq}_i \lbrace 1, v_{i\alpha} \rbrace} = 0$
which yield the solvability conditions
\begin{align}
\label{eq:solvability_f}
\integral{f_i^{(1)} \lbrace 1, v_{i\alpha} \rbrace} &= \integral{f_i^{(2)} \lbrace 1, v_{i\alpha} \rbrace} = \cdots = 0, \\
\label{eq:solvability_h}
\integral{h_i^{(1)} \lbrace 1, v_{i\alpha} \rbrace} &= \integral{h_i^{(2)} \lbrace 1, v_{i\alpha} \rbrace} = \cdots = 0, 
\end{align}
using~eqs.~(\ref{eq:expand_f}) to~(\ref{eq:expand_h}) and (\ref{eq:f_h_zeroth_order}).

The terms of order $\epsilon^1$ lead to
\begin{equation}
\label{eq:first_order}
(\partial_t^{(1)} + \partial_\alpha v_{i\alpha}) \feq_i = - 2\beta\left[f_i^{(1)} + \tfrac{1}{2}(\gamma-2) h_i^{(1)} \right],
\end{equation}
from which we can recover the hydrodynamic equations of mass and momentum to first order by taking the zeroth and first order moment of \eqref{eq:first_order} and using condititions~\eqref{eq:solvability_f} and~\eqref{eq:solvability_h} and the definition of $\Pi_{\alpha\beta}^{\rm eq}$
\begin{align}
\label{eq:continuity_first_order}
\partial_t^{(1)} \rho &= - \partial_\alpha (\rho u_\alpha) \\
\label{eq:momentum_first_order}
\partial_t^{(1)} u_\alpha &= \tfrac{1}{\rho}u_\alpha\partial_\beta(\rho u_\beta) - \tfrac{1}{\rho}\partial_\beta \Pi_{\alpha\beta}^{\rm eq}.
\end{align}

Collecting terms of order $\epsilon^2$ results in
\begin{equation}
\label{eq:second_order}
\left[ \partial_t^{(2)} + \tfrac{1}{2}(\partial_t^{(1)} + \partial_\alpha v_{i\alpha})(\partial_t^{(1)} + \partial_\beta v_{i\beta}) \right] \feq_i +  (\partial_t^{(1)} + \partial_\alpha v_{i\alpha})f_i^{(1)} = -2\beta\left[ f_i^{(2)} + \tfrac{1}{2}(\gamma-2)h_i^{(2)} \right].
\end{equation}
From eq.~\eqref{eq:first_order} we get the relation
\begin{equation}
\label{eq:f1}
f_i^{(1)} = -\tfrac{1}{2\beta}(\partial_t^{(1)} + \partial_\alpha v_{i\alpha}) \feq_i + \tfrac{1}{2}(2-\gamma)h^{(1)}_i,
\end{equation}
which, once inserted in eq.~\eqref{eq:second_order}, leads to 
\begin{equation}
\label{eq:second_order2}
\begin{split}
\left[ \partial_t^{(2)} + (\tfrac{1}{2}-\tfrac{1}{2\beta})(\partial_t^{(1)} + \partial_\alpha v_{i\alpha})(\partial_t^{(1)} + \partial_\beta v_{i\beta}) \right] \feq_i +  \tfrac{1}{2}(\partial_t^{(1)} + \partial_\alpha v_{i\alpha})(2-\gamma)h^{(1)}_i \\
= -2\beta\left[ f_i^{(2)} + \tfrac{1}{2}(\gamma-2)h_i^{(2)} \right].
\end{split}
\end{equation}

Taking the zeroth order moment thereof and making use of
eqs.~(\ref{eq:solvability_f}), (\ref{eq:solvability_h}), (\ref{eq:continuity_first_order}) and (\ref{eq:momentum_first_order})
yields the vanishing second order contribution to the continuity equation
\begin{equation}
\label{eq:continuity_second_order}
\partial_t^{(2)} \rho = 0.
\end{equation}
The first order moment, however, will render different outcomes for the four KBC models depending on the second order moment of $h$,
\begin{equation}
\label{eq:momentum_second_order}
\partial_t^{(2)} u_\alpha = \tfrac{1}{\rho} \left( \tfrac{1}{2\beta} -\tfrac{1}{2} \right) \partial_\beta\left[  \partial_t^{(1)} \Pi_{\alpha\beta}^{\rm eq} + \partial_\mu Q_{\alpha\beta\mu}^{\rm eq} \right] + \tfrac{1}{\rho}\partial_\beta\left[ (\gamma-2) \tfrac{1}{2} \integral{h_i^{(1)} v_{i\alpha} v_{i\beta} } \right].
\end{equation}
For models A and C the last term on the right hand side vanishes according to table \ref{tab:sh_moments2} and we get
\begin{equation}
\label{eq:momentum_second_order_AC}
\partial_t^{(2)} u_\alpha = \tfrac{1}{\rho} \partial_\beta\left[  \left(\tfrac{1}{2\beta} - \tfrac{1}{2} \right) \left(\partial_t^{(1)} \Pi_{\alpha\beta}^{\rm eq} + \partial_\mu Q_{\alpha\beta\mu}^{\rm eq}\right) \right],
\end{equation}
while for models B and D one must first analyze the second moment of $h$. 
As only the trace $\integral{h_i^{(1)} v_{i\alpha} v_{i\alpha} }$ will contribute we can get the following relation using eq.~\eqref{eq:f1}, 
condition $f_i^{(1)} = s_i^{(1)} + h_i^{(1)}$ and table~\ref{tab:sh_moments2}
\begin{equation}
\label{eq:second_order_h}
\begin{split}
\integral{h_i^{(1)} v_{i\alpha} v_{i\beta} } = \tfrac{1}{2}\integral{h_i^{(1)} v_{i\mu} v_{i\mu} }\delta_{\alpha\beta} 
= -\integral{\tfrac{1}{\gamma}\left(s_i^{(1)} + \tfrac{1}{2\beta}(\partial_t^{(1)}+\partial_\sigma v_{i\sigma})\feq_i\right)v_{i\mu} v_{i\mu} }\delta_{\alpha\beta} \\
= -\tfrac{1}{2\gamma\beta}\left( \partial_t^{(1)}\Pi_{\mu\mu}^{\rm eq} + \partial_\sigma Q_{\sigma\mu\mu}^{\rm eq} \right)\delta_{\alpha\beta} .
\end{split}
\end{equation}
When inserted in eq.~\eqref{eq:continuity_second_order}, this gives the second order contribution to the momentum equation
\begin{equation}
\begin{split}
\label{eq:momentum_second_order_BD}
\partial_t^{(2)} u_\alpha = \tfrac{1}{\rho}  \partial_\beta\left[  \left( \tfrac{1}{2\beta} -\tfrac{1}{2} \right) \left( \partial_t^{(1)} \Pi_{\alpha\beta}^{\rm eq} + \partial_\mu Q_{\alpha\beta\mu}^{\rm eq} -\tfrac{1}{2}\left( \partial_t^{(1)}\Pi_{\mu\mu}^{\rm eq} + \partial_\sigma Q_{\sigma\mu\mu}^{\rm eq} \right)\delta_{\alpha\beta} \right) \right] \\
+ \tfrac{1}{\rho}\partial_\alpha\left[ ( \tfrac{1}{\gamma\beta}-\tfrac{1}{2})\tfrac{1}{2} \left( \partial_t^{(1)}\Pi_{\mu\mu}^{\rm eq} + \partial_\sigma Q_{\sigma\mu\mu}^{\rm eq} \right)\right].
\end{split}
\end{equation}

For completeness let us now compute the first order terms of the pressure tensor for models A, C and B, D
which yield, by substituting the equilibrium values eqs.~(\ref{eq:Peq}) and (\ref{eq:Qeq})
and using the first order results 
eqs.~(\ref{eq:continuity_first_order}) and (\ref{eq:momentum_first_order})
%
\begin{equation}
\label{eq:P1AC}
\Pi_{\alpha\beta}^{(1)} = -\tfrac{\rho c_s^2}{2\beta} \left[ \partial_\alpha u_\beta + \partial_\beta u_\alpha \right],
\end{equation}
and
\begin{equation}
\label{eq:P1BD}
\Pi_{\alpha\beta}^{(1)} = -\tfrac{\rho c_s^2}{2\beta} \left[ \partial_\alpha u_\beta + \partial_\beta u_\alpha - \tfrac{2}{D}\partial_\mu u_\mu \delta_{\alpha\beta} \right] - \tfrac{\rho c_s^2}{\gamma\beta}\left[ \tfrac{2}{D}\partial_\mu u_\mu \delta_{\alpha\beta} \right],
\end{equation}
respectively.

By summing up the contributions from first and second order in $\epsilon$, using the
eqs.~(\ref{eq:Peq}), (\ref{eq:Qeq}), (\ref{eq:continuity_first_order}), (\ref{eq:momentum_first_order}) and (\ref{eq:continuity_second_order})
and reintroducing dimensional variables we recover the isothermal Navier-Stokes equations at reference temperature $T_0=c_s^2$
\begin{align}
\label{eq:continuity}
\partial_t \rho &= - \partial_\alpha(\rho u_\alpha), \\
\label{eq:momentum}
\partial_t u_\alpha &=  -u_\beta\partial_\beta u_\alpha - \tfrac{1}{\rho}\partial_\alpha(c_s^2\rho)
+ \tfrac{1}{\rho}\partial_\beta\left[ \nu \rho \left( \partial_\alpha u_\beta + \partial_\beta u_\alpha - \tfrac{2}{D}\partial_\mu u_\mu \delta_{\alpha\beta} \right) \right]
+ \tfrac{2}{D\rho}\partial_\alpha\left[ \xi\rho\partial_\mu u_\mu \right],
\end{align}
with the kinematic (shear) and bulk viscosity coefficients
\begin{equation}
\label{eq:viscosity2}
\nu = c_s^2\left( \tfrac{1}{2\beta} - \tfrac{1}{2} \right), ~ \xi =  \begin{cases} ~\nu & \mbox{models A and C } \\
c_s^2\left( \tfrac{1}{\gamma\beta} - \tfrac{1}{2} \right) & \mbox{models  B and D}. \end{cases}
\end{equation}

Note that for the quasi-incompressible LB method the bulk viscosity term is small and can be considered as an artefact of the numerical method. We also point out that the bulk viscosity $\xi$ for KBC B and D depends on time and space along with the stabilizer $\gamma$.

\begin{figure}[ht]
    \centering
    \begin{subfigure}[b]{0.49\textwidth}
    		\centering
    		\includegraphics[width=\textwidth]{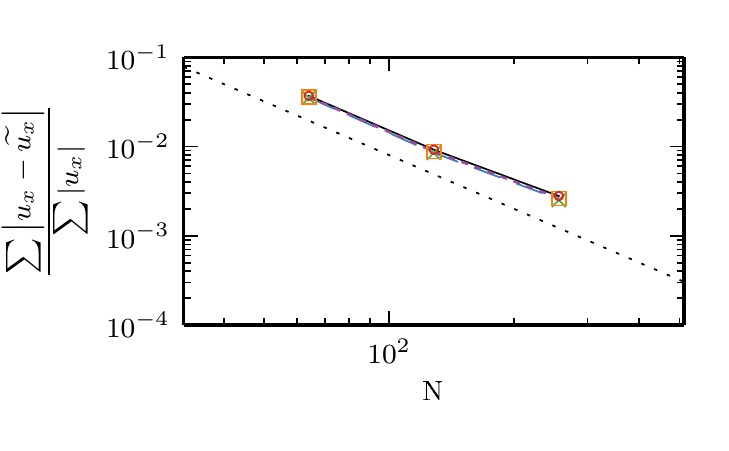}\llap{\parbox[b]{2.1in}{\textbf{(a)}\\\rule{0ex}{0.55in}}}
    	\end{subfigure}
    	\begin{subfigure}[b]{0.49\textwidth}
    		\centering
    		\includegraphics[width=\textwidth]{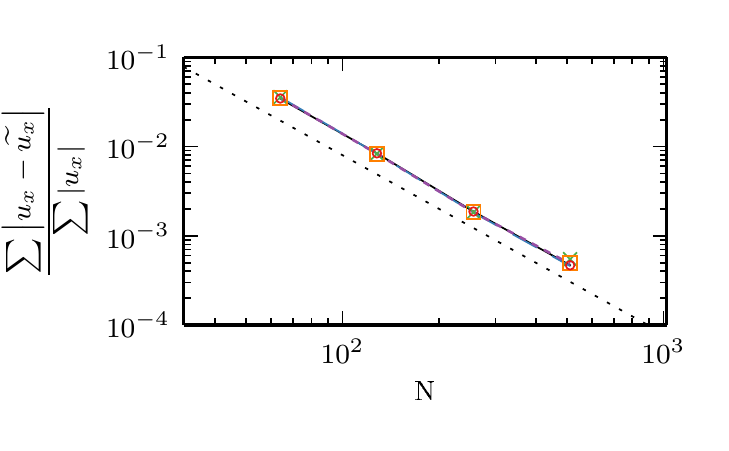}\llap{\parbox[b]{2.1in}{\textbf{(b)}\\\rule{0ex}{0.55in}}}
    	\end{subfigure}
	\caption{Convergence rate for Green-Taylor vortex at $t=t_c$. (a) $Re = 100, u_0=0.03$, (b) $Re=1000, u_0=0.04$. LBGK (\lbgksym), KBC A (\kbcAsym), KBC B (\kbcBsym), KBC C (\kbcCsym), KBC D (\kbcDsym), ELBM (\elbmsym), second order convergence (fine dotted).}
	\label{fig:GTconvergence}
\end{figure}

\section{Green-Taylor Vortex Flow}

For all of the flows considered in this paper periodic boundary conditions are applied in order to separate the accuracy of the scheme from influences of boundary conditions. The KBC scheme is validated for the Green-Taylor vortex flow in a first numerical example. The analytical solution for the Green-Taylor vortex flow is given by $\bm{u}(x,y,t) = \bm{\nabla} \times \left[(\bm{u}_0/k_2)\cos(k_1 x) \cos(k_2 y) \exp(-\nu(k_1^2 + k_2^2)t)\right]$
where we have chosen $k_1 = 1$, $k_2 = 4$ and the pressure $p_0 = \rho c_s^2$ is initialized using $\rho_0 = 1$.
The populations were initialized using Grad's approximation \cite{Karlin} 
\begin{equation}
\label{eq:Grad}
f_i^\ast(\rho, \bm{u}, \bm{\Pi}) = W_i\left[ \rho + \tfrac{\rho u_\alpha v_{i\alpha}}{c_s^2} 
+ \tfrac{1}{2c_s^4}(\Pi_{\alpha\beta} - \rho c_s^2 \delta_{\alpha\beta})(v_{i\alpha}v_{i\beta} - c_s^2\delta_{\alpha\beta}) \right]
\end{equation}
where the pressure tensor was taken in the form $\Pi_{\alpha\beta}^{\rm eq} + \Pi_{\alpha\beta}^{(1)}$ 
(see eqs.~(\ref{eq:Peq}) and (\ref{eq:P1AC})).
The domain is confined in $0<x,y<2\pi$ covered by a mesh of $N \times N$ lattice nodes, the Reynolds number is defined as $Re = u_0 N/\nu$ and the decay half-time of the flow is given by $t_c = \ln 2/\left[\nu(k_1^2 + k_2^2)\right]$ lattice time steps.

The Reynolds number was set to ${\rm Re}=100$ in a first experiment and ${\rm Re}=1000$ in a second simulation while the initial velocity was $u_0=0.03$ and $u_0=0.04$, respectively. Resolution $N$ was varied between $\lbrace64,128,256\rbrace$ in the former and between $\lbrace62,128,256,512\rbrace$ in the latter case. KBC models A-D, LBGK and ELBM were run and compared to the analytic solution. Figure \ref{fig:GTconvergence} shows the convergence for the relative error at time $t=t_c$. Second order rate is clearly observed for all models, moreover, all the models are performing almost identically.

\begin{figure}[ht]
	\centering
	\includegraphics[width=0.9\textwidth]{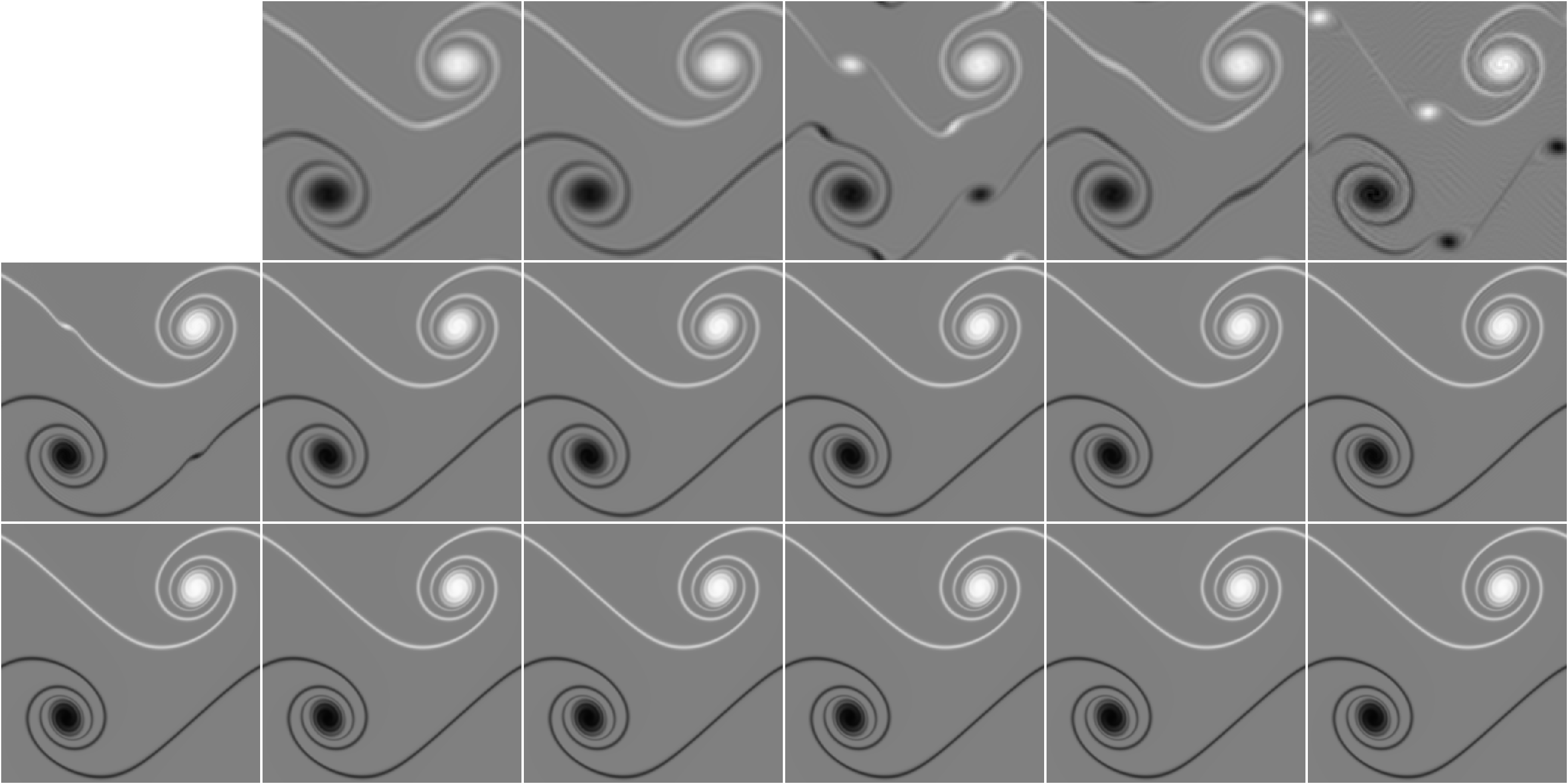}
	\caption{Vorticity field at $t=t_c$ for $\kappa=80$, $u_0=0.04$. Columns: LBGK, KBC A, KBC B, KBC C, KBC D, ELBM, respectively, 
	rows: Resolution $N = 128, 256, 512$.}
	\label{fig:SL_vort_4}
\end{figure}

\section{Doubly Periodic Shear Layer}

To probe the KBC models for their performance in under-resolved simulations of smooth flows with sharp features the doubly periodic double shear layer with a slight perturbation studied extensively in \cite{Minion97} was used as a benchmark. Initial conditions are given by
\vspace{-0.2cm}
\begin{align}
\begin{split}
u_x &= \left\{ \begin{array}{ll}
                             u_0 \tanh\left(\kappa \left(y/N-0.25\right) \right), y \leq N/2, \\
                             u_0 \tanh\left(\kappa \left(0.75-y/N\right) \right), y > N/2,
                           \end{array}\right. \nonumber\\
u_y &= \delta u_0 \sin\left(2\pi \left(x/N + 0.25 \right) \right).
\end{split}
\end{align}
Here $N$ is the number of grid points in both $x$ and $y$ directions while periodic boundary conditions are applied in both directions. Grad's approximation \eqref{eq:Grad} was used to initialize the flow field while initial density was set to unity.
The parameter $\kappa$ controls the width of the shear layer while $\delta$ is a small perturbation of the velocity in $y$-direction which initiates a Kelvin-Helmholtz instability causing the roll up of the anti-parallel shear layers. $u_0$ is the initial magnitude of the $x$-velocity while the Reynolds number is defined as ${\rm Re} = u_0 N/\nu$ and the turnover time is $t_c = N/u_0$ lattice time steps. 

In \cite{Minion97} it is demonstrated that all the numerical methods investigated therein produce spurious additional vortex roll-ups as a consequence of under-resolution. Effectively no convergence could be reported until the resolution was sufficiently high for the additional vortices to disappear.

\begin{figure}[!h]
    \centering
    \begin{subfigure}[b]{0.45\textwidth}
    		\centering
    		\includegraphics[width=\textwidth]{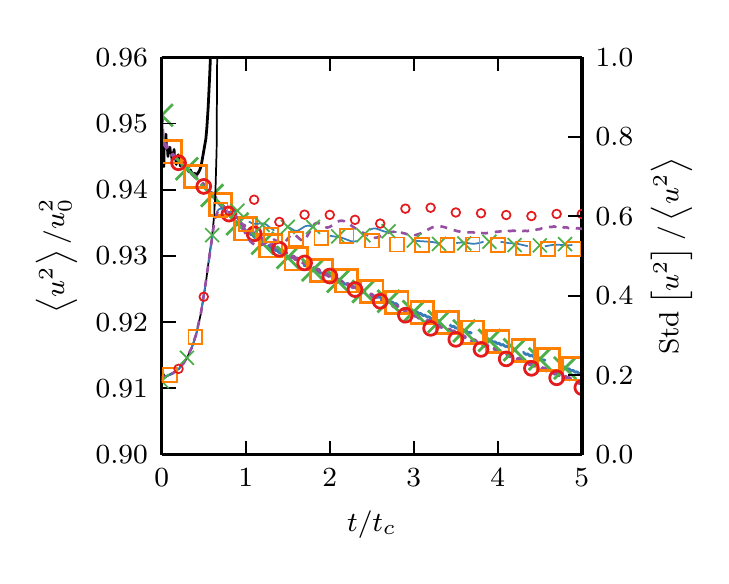}\llap{\parbox[b]{2.0in}{\textbf{(a)}\\\rule{0ex}{0.5in}}}
    	\end{subfigure}~
    	\begin{subfigure}[b]{0.45\textwidth}
    		\centering
    		\includegraphics[width=\textwidth]{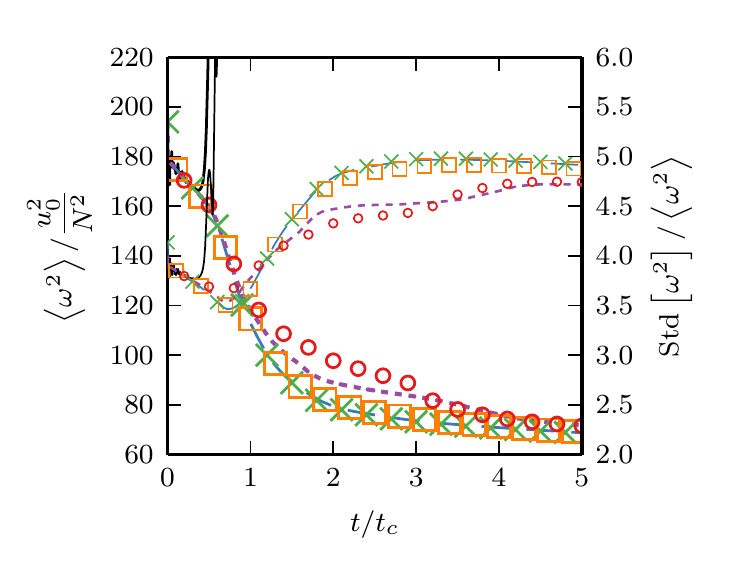}\llap{\parbox[b]{2.0in}{\textbf{(b)}\\\rule{0ex}{0.5in}}}
    	\end{subfigure}
    	\vspace*{-0.5cm}
    	
    	\begin{subfigure}[b]{0.45\textwidth}
    		\centering
    		\includegraphics[width=\textwidth]{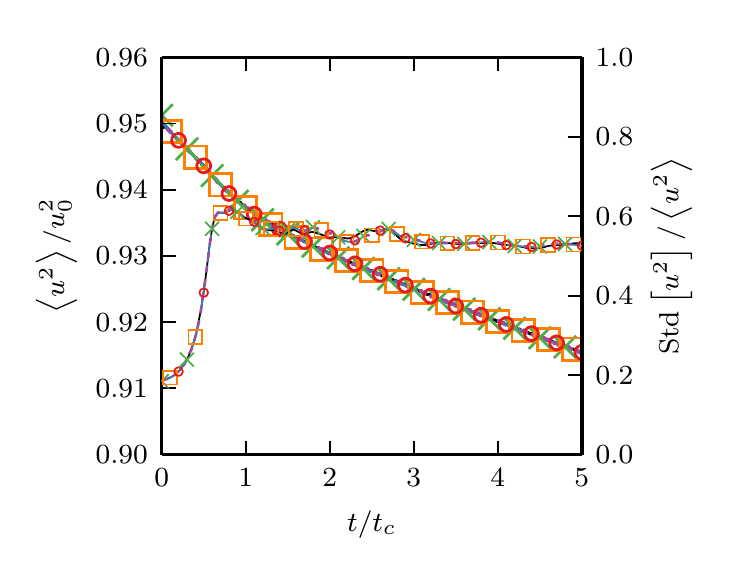}\llap{\parbox[b]{2.0in}{\textbf{(c)}\\\rule{0ex}{0.5in}}}
    	\end{subfigure}~
    	\begin{subfigure}[b]{0.45\textwidth}
    		\centering
    		\includegraphics[width=\textwidth]{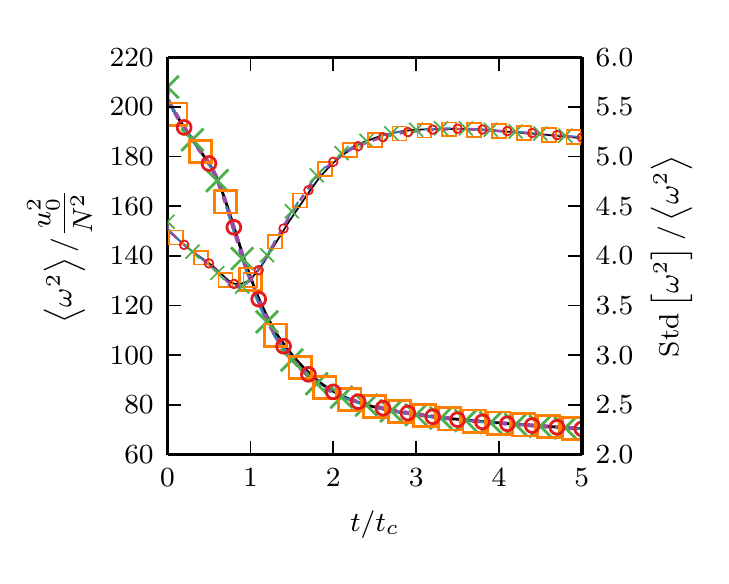}\llap{\parbox[b]{2.0in}{\textbf{(d)}\\\rule{0ex}{0.5in}}}
    	\end{subfigure}
    	\vspace*{-0.4cm}
    		
    	\caption{Evolution of kinetic energy and enstrophy (mean: left axis and large symbols, standard deviation: right axis and smaller symbols). Resolution: a) and b) $N=128$, c) and d) $N=256$, respectively. LBGK (\lbgksym), KBC A (\kbcAsym), KBC B (\kbcBsym), KBC C (\kbcCsym), KBC D (\kbcDsym), ELBM (\elbmsym).
    	Results represented by symbols have been subsampled for clarity.}
    	\label{fig:SL_energy}
\end{figure}

We pose the question whether the different relaxation for the non-hydrodynamic higher order moments are advantageous for the performance in {under-resolved} cases. To this end let us consider a thin shear layer case with $\kappa = 80$, ${\rm Re} = 30'000$ and $u_0 = 0.04$. We compare KBC models A-D, LBGK and ELBM. Figure \ref{fig:SL_vort_4} shows the vorticity field at $t=t_c$ for the six schemes under consideration for different resolutions $N=\lbrace128, 256, 512\rbrace$. LBGK becomes unstable even before $t=t_c$ is reached (see also fig. \ref{fig:SL_energy} a) and b)) for $N=128$. 
For $N=128$, model C and ELBM clearly show formation of additional roll-ups whereas model D produces comparatively small instabilities. Models A and B (central moments) capture the flow features quite accurately while model B seems to perform slightly better. 
For the next higher resolution under consideration, $N=256$, LBGK survives but features two small additional vortices while the other models capture the main flow features well. For the largest resolution, $N=512$, the models are essentially indistinguishable. 

In summary, KBC B performs qualitatively better in the under-resolved situation than the other models while for the still slightly under-resolved case, $N=256$, the KBC models and ELBM give comparably good results while LBGK still features spurious vortices at this resolution.

\begin{figure}[!h]
    \centering
    \includegraphics[width=0.49\textwidth]{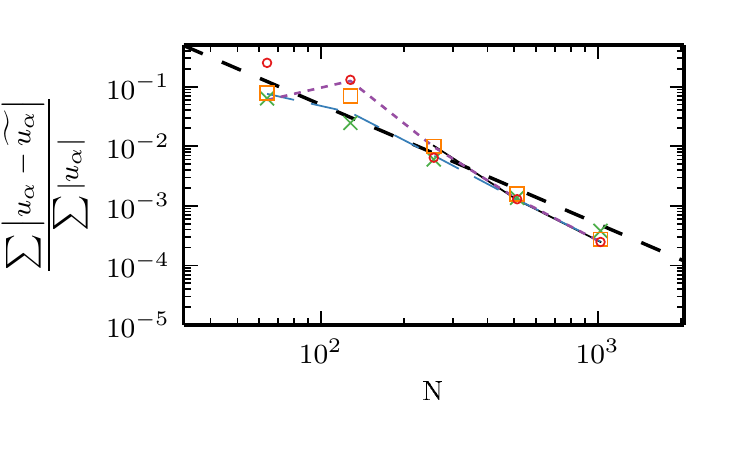}
    \caption{Convergence rate for doubly periodic shear layer at $t=t_c$ and $Re = 30000, \kappa = 80, u_0 = 0.04$. LBGK (\lbgksym), KBC A (\kbcAsym), KBC B (\kbcBsym), KBC C (\kbcCsym), KBC D (\kbcDsym), ELBM (\elbmsym), second order convergence (fine dotted). Error with respect to reference solution (LBGK at $N=2048$ resolution).}
    \label{fig:SL_convergence}
\end{figure}

These findings are also reflected in figure \ref{fig:SL_convergence} where the second order convergence is reached for all models after $N=256$. KBC models B and A clearly outperform the other schemes in the under-resolved cases.

Let us now consider the energy and enstrophy decay, eq.~\eqref{fig:SL_energy}, where we report both the mean and the fluctuations (RMS) over time. We first remark that the methods converge to each other for $N=256$ while there is also evidence that the simulation is resolved as the statistics do not change for the next higher resolution, $N=512$. The energy decay is rather similar for all the models across all resolutions, indicated by both mean and standard deviation, except for ELBM which shows slightly different results for the mean and fluctuations at $N\leq128$. The mean enstrophy and fluctuations over time are clearly better captured for the KBC models A, B and D in the under-resolved situation compared to ELBM and KBC C. This is also in accordance with the visual impression of the vorticity structure (fig. \ref{fig:SL_vort_4}). 
In summary, the low order statistics seem not affected by the KBC treatment of the higher order moments.

\section{Decaying Two-Dimensional Turbulence}

The third and final numerical example considered in this paper is the simulation of a turbulent, albeit two-dimensional, flow starting from a random initial condition and decaying with time. 

Decaying two-dimensional turbulence is characterized by the formation of vortices in the early stage (vortex generation period) which leads to spatially separated coherent structures (which account for the vorticity extrema) which typically have long lifetimes compared to the eddy turnover time and undergo passive advection and vortex-vortex interaction \cite{Mcwilliams2dVortices}. These vortices can persist, grow over time as they merge with weaker structures of same-sign vorticity and influence the whole field \cite{bracco2000velocity}. A large amount of the enstrophy is concentrated within the large-scale vortices that decay slower than the background vorticity field between the vortices \cite{SantangeloBenzi}. The total energy is roughly constant while enstrophy is decaying. According to \cite{Kraich2d} the enstrophy follows a direct cascade from large to small scales, much alike the energy cascade in three-dimensional turbulence, while the energy shows an inverse cascade from small to large scales. Classical Kolmogorov-Batchelor scaling theory predicts a slope of $k^{-3}$ and $k^{-1}$ for the energy ($E$) and enstrophy ($Z$) spectrum, respectively, where $k$ is the wave-vector magnitude.

\begin{figure}[t]
    \centering
	\includegraphics[width=\textwidth]{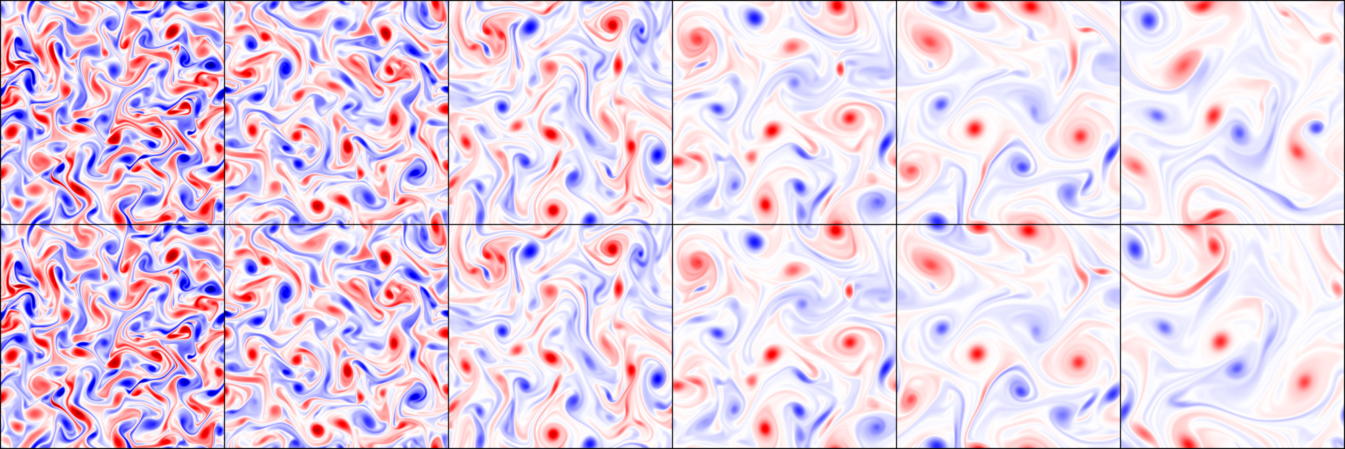}
	\caption{Vorticity field for decaying two-dimensional turbulence ${\rm Re}=13'134$, $N=1024$, for LBGK (first row) and KBC B (second row) and times $t/t_e=\lbrace10,20,40,60,80,100\rbrace$ from left to right.}
	\label{fig:2dturbvortR13}
\end{figure}

In general, for the simulation of fluid turbulence it useful to study the following questions:
\begin{enumerate}
\item is the dynamics of a fully developed turbulent flow accurately captured (i.e. initial vortex formation, emergence of coherent structures, vortex-vortex interactions, decay of vortex density)?
\item are near {grid-scale} structures with large gradients (small vortices) well represented for sufficiently high Reynolds numbers?
\item is the stability affected by the these {lattice-scale} structures? 
\item are low-order statistics well represented and is the numerical scheme correctly modelling the physical dissipation (i.e. decay of enstrophy, scaling laws)?
\item how good is the performance for very large Reynolds numbers in an under-resolved simulation?
\end{enumerate}

The initial conditions for all subsequent simulations are given by constructing a zero-mean Gaussian random field in Fourier-space with random Fourier-phases and amplitudes proportional to the prescribed spectral density of the stream function $\Psi(k) = k^{-4} Z(k) = k^{-2} E(k)$ from which an incompressible velocity field for a domain of $N \times N$ lattice nodes is obtained. The energy spectral density function is given by
\begin{equation}
E(k)=C_0 k^A\left[1 + (k/k_0)^{B+1} \right]^{-1}
\end{equation}
where $C_0$ is a normalization constant and the parameters $A=6$ and $B=17$ such that the energy spectrum is narrow banded and reasonably peaked at small wave numbers \cite{Millen}.

\begin{table}[t]
\centering
\scalebox{0.75}{
\begin{tabular}{cccccccccc}
\toprule
 & \multicolumn{3}{c}{${\rm Re} = 13'134$} & \multicolumn{3}{c}{${\rm Re} = 1.5\cdot 10^5$} & \multicolumn{3}{c}{${\rm Re} = 1.6\cdot 10^6$}\\
\cmidrule(rl){2-4} \cmidrule(rl){5-7} \cmidrule(rl){8-10}
$N$ & $256$ & $512$ & $1024$ & $1024$ & $2048$ & $4096$ & $1024$ & $2048$ & $4096$ \\
\midrule
$\left\langle E(0) \right\rangle$ & $1.645\cdot 10^{-4}$ & $1.645\cdot 10^{-4}$ & $1.645\cdot 10^{-4}$ & $1.209\cdot 10^{-4}$ & $1.209\cdot 10^{-4}$ & $1.209\cdot 10^{-4}$ & $1.209\cdot 10^{-4}$ & $1.209\cdot 10^{-4}$ & $1.209\cdot 10^{-4}$ \\
$\left\langle Z(0) \right\rangle$ & $8.455\cdot 10^{-6}$ & $2.137\cdot 10^{-6}$ & $5.358\cdot 10^{-7}$ & $3.815\cdot 10^{-6}$ & $9.602\cdot 10^{-7}$ & $2.404\cdot 10^{-7}$ & $3.815\cdot 10^{-6}$ & $9.602\cdot 10^{-7}$ & $2.404\cdot 10^{-7}$ \\
$\nu$                             & $3.533\cdot 10^{-4}$ & $7.066\cdot 10^{-4}$ & $1.413\cdot 10^{-3}$ & $1.062\cdot 10^{-4}$ & $2.123\cdot 10^{-4}$ & $4.246\cdot 10^{-4}$ & $9.952\cdot 10^{-6}$ & $1.990\cdot 10^{-5}$ & $3.981\cdot 10^{-5}$ \\
$t_e$                             & $344$ & $684$ & $1366$ & $512$ & $1021$ & $2039$ & $512$ & $1021$ & $2039$ \\
\bottomrule
\end{tabular}
}
\caption{Characteristics for two-dimensional turbulence simulations (lattice units).}
\label{tab:2dturbsetup}
\end{table}

\begin{figure}[b]
    \centering
    \begin{subfigure}[b]{0.43\textwidth}
    		\centering
    		\includegraphics[width=\textwidth]{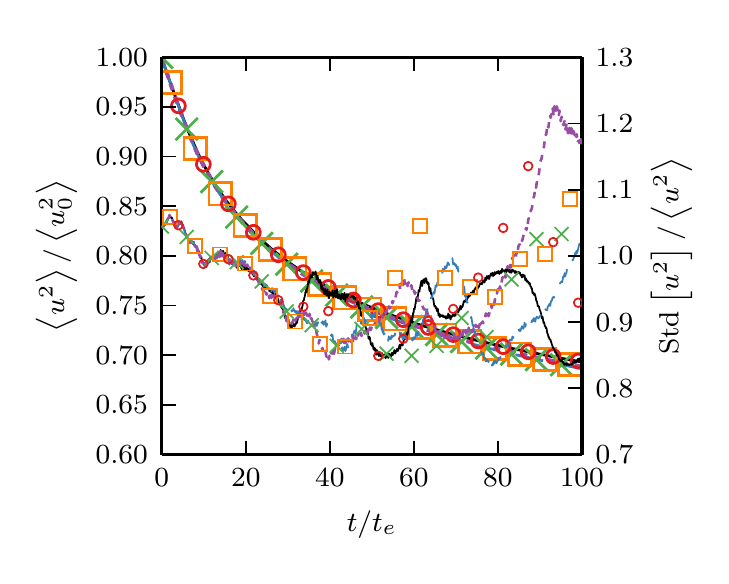}\llap{\parbox[b]{1.9in}{\textbf{(a)}\\\rule{0ex}{0.5in}}}
    	\end{subfigure}
    	\begin{subfigure}[b]{0.43\textwidth}
    		\centering
    		\includegraphics[width=\textwidth]{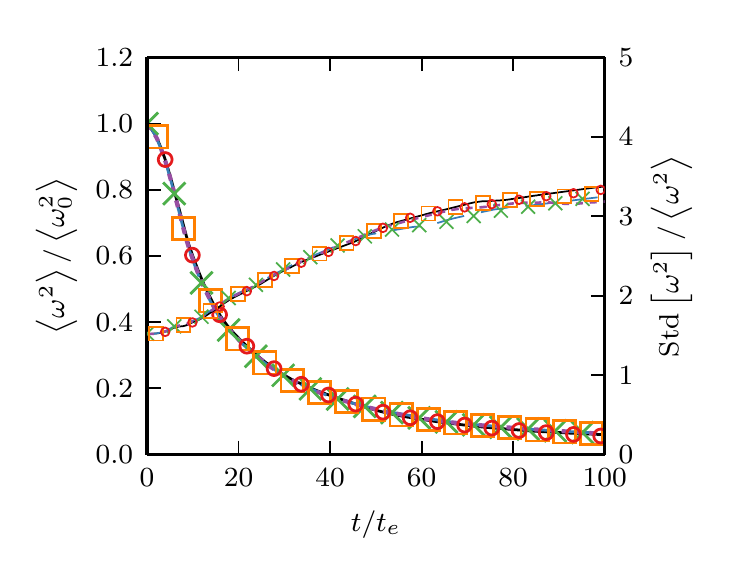}\llap{\parbox[b]{2.0in}{\textbf{(b)}\\\rule{0ex}{0.5in}}}
    	\end{subfigure}
    	\vspace*{-0.3cm}
    	
    	\begin{subfigure}[b]{0.43\textwidth}
    		\centering
    		\includegraphics[width=\textwidth]{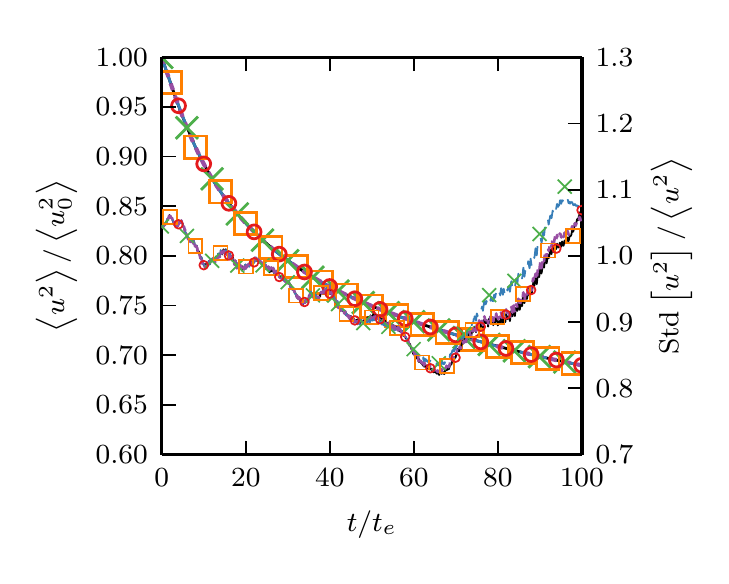}\llap{\parbox[b]{1.9in}{\textbf{(c)}\\\rule{0ex}{0.5in}}}
    	\end{subfigure}
    	\begin{subfigure}[b]{0.43\textwidth}
    		\centering
    		\includegraphics[width=\textwidth]{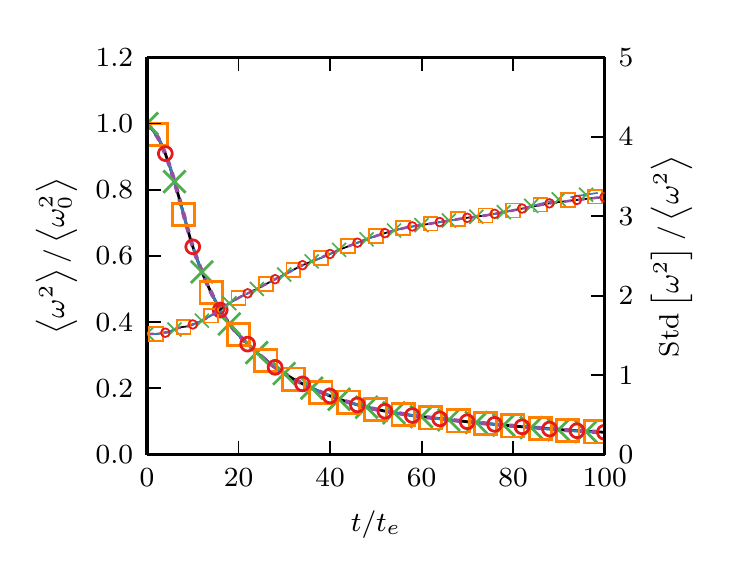}\llap{\parbox[b]{2.0in}{\textbf{(d)}\\\rule{0ex}{0.5in}}}
    	\end{subfigure}
	\vspace*{-0.3cm}
    	
    	\caption{Evolution of kinetic energy and enstrophy for $Re = 13'134$. Mean: left axis and large symbols, standard deviation: right axis and smaller symbols. Resolution $N=256, 1024$ from top to bottom, respectively. LBGK (\lbgksym), KBC A (\kbcAsym), KBC B (\kbcBsym), KBC C (\kbcCsym), KBC D (\kbcDsym), ELBM (\elbmsym). Results represented by symbols have been subsampled for clarity.}
    	\label{fig:2dturbenergyR13}
\end{figure}

\begin{figure}[t]
    \centering
    \begin{subfigure}[b]{0.45\textwidth}
    		\centering
   		\includegraphics[width=\textwidth]{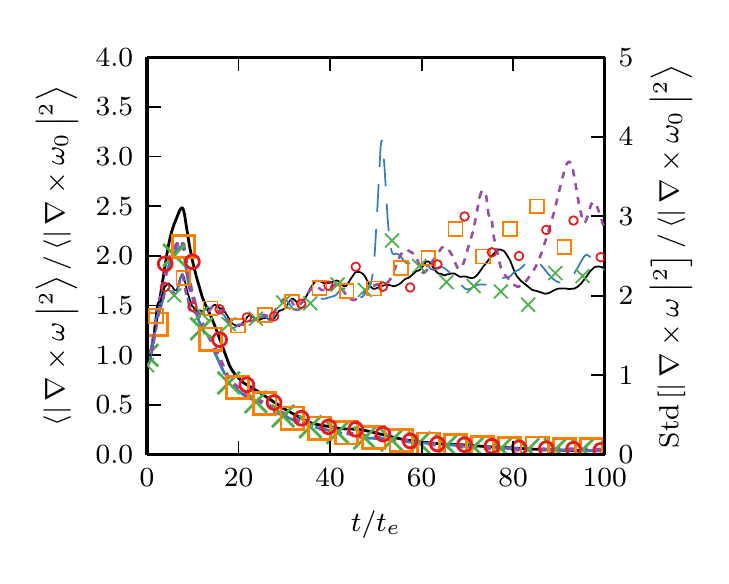}\llap{\parbox[b]{2.05in}{\textbf{(a)}\\\rule{0ex}{0.5in}}}
    	\end{subfigure}
    	\begin{subfigure}[b]{0.45\textwidth}
    		\centering
    		\includegraphics[width=\textwidth]{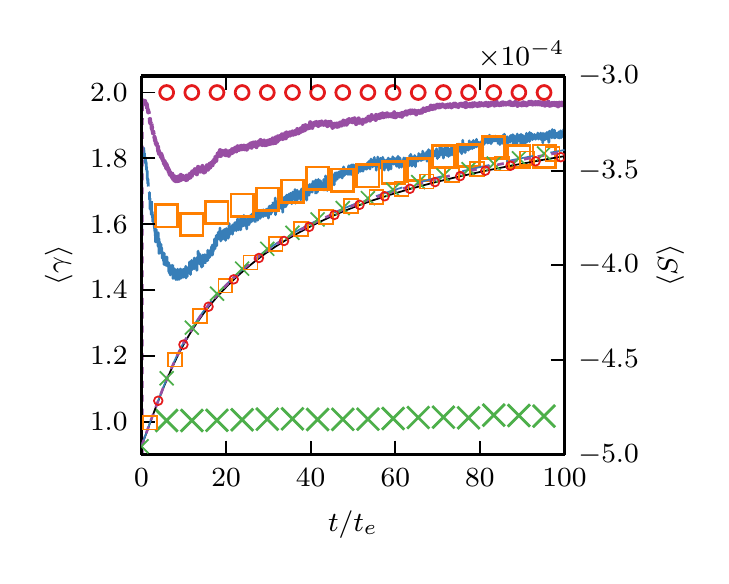}\llap{\parbox[b]{2.05in}{\textbf{(b)}\\\rule{0ex}{0.5in}}}
    	\end{subfigure}
	\vspace*{-0.4cm}
    		
    	\begin{subfigure}[b]{0.45\textwidth}
    		\centering
   		\includegraphics[width=\textwidth]{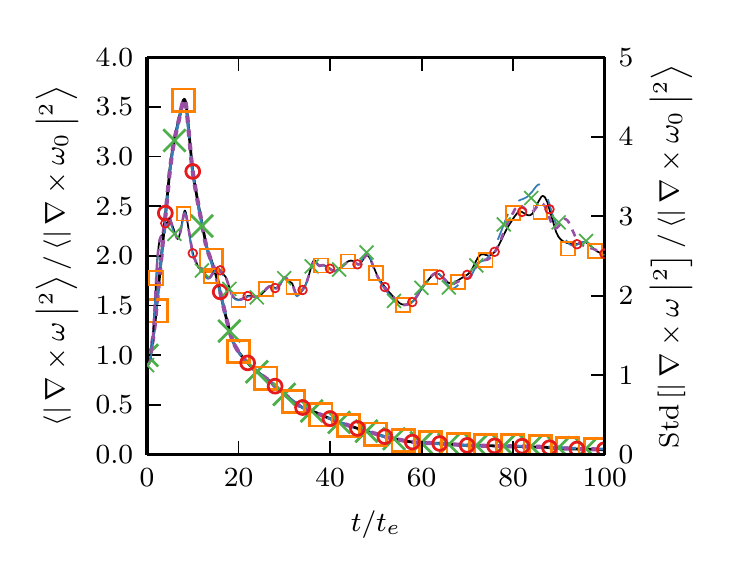}\llap{\parbox[b]{2.05in}{\textbf{(c)}\\\rule{0ex}{0.5in}}}
    	\end{subfigure}
    	\begin{subfigure}[b]{0.45\textwidth}
    		\centering
    		\includegraphics[width=\textwidth]{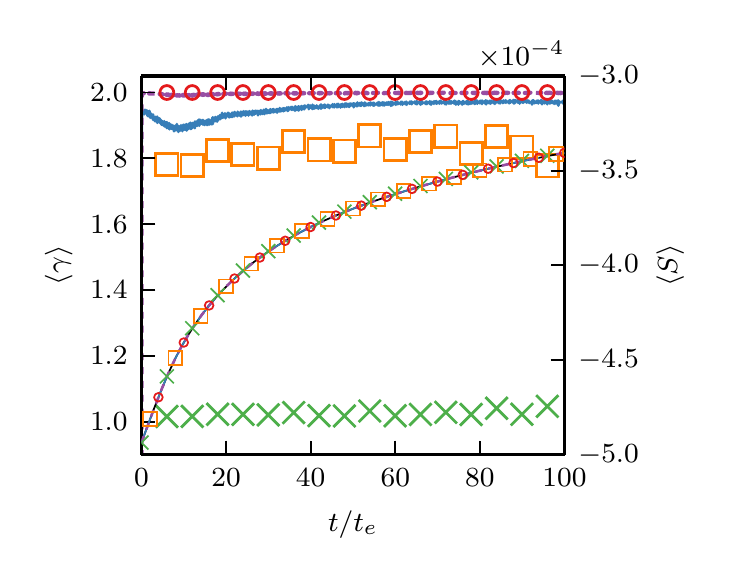}\llap{\parbox[b]{2.05in}{\textbf{(d)}\\\rule{0ex}{0.5in}}}
    	\end{subfigure}
	\vspace*{-0.3cm}
    	
     	\caption{Evolution of mean stabilizer $\gamma$, mean entropy (right column, left y-axis, large symbols and right y-axis, small symbols, respectively) and  palinstrophy (left column, mean: left y-axis, large symbols, standard deviation: right y-axis, small symbols) for $Re = 13'134$. Resolution $N=256, 1024$ from top to bottom, respectively. LBGK (\lbgksym), KBC A (\kbcAsym), KBC B (\kbcBsym), KBC C (\kbcCsym), KBC D (\kbcDsym), ELBM (\elbmsym). Results represented by symbols have been subsampled for clarity.}
     	\label{fig:2dturbPalinGammaR13}
\end{figure}

\begin{figure}[!h]
    \centering

    	\begin{subfigure}[b]{0.45\textwidth}
    		\centering
   		\includegraphics[width=\textwidth]{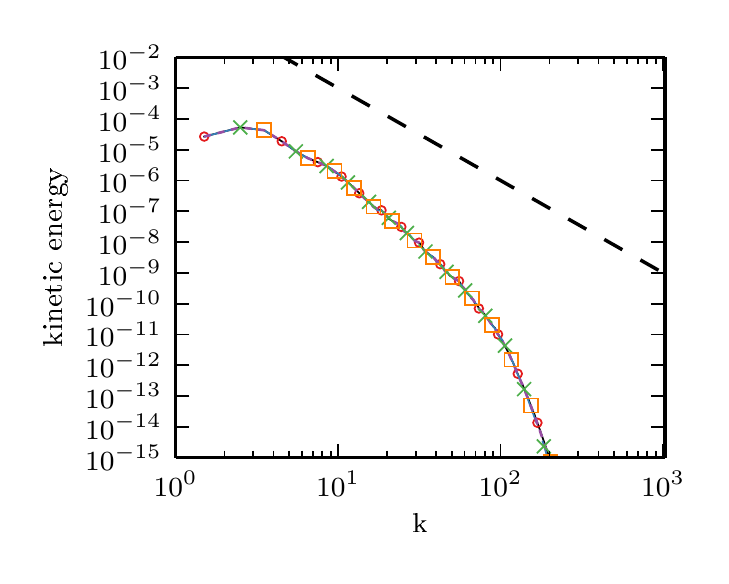}\llap{\parbox[b]{1.95in}{\textbf{(a)}\\\rule{0ex}{0.5in}}}
    	\end{subfigure}
    	\begin{subfigure}[b]{0.45\textwidth}
    		\centering
    		\includegraphics[width=\textwidth]{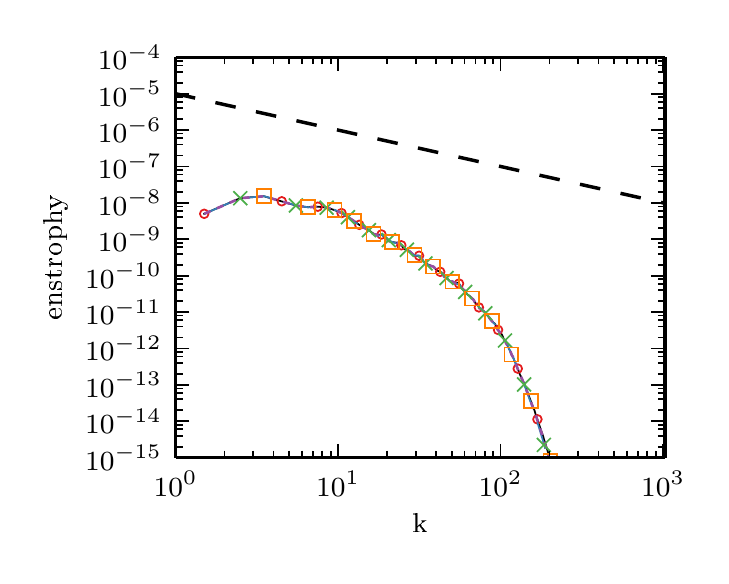}\llap{\parbox[b]{1.95in}{\textbf{(b)}\\\rule{0ex}{0.5in}}}
    	\end{subfigure}
     \vspace*{-0.3cm}

     	\caption{Energy (a) and enstrophy spectra (b) for two-dimensional turbulence at ${\rm Re} = 13'134$ for $N=1024$ at time $t/t_e=50$. LBGK (\lbgksym), KBC A (\kbcAsym), KBC B (\kbcBsym), KBC C (\kbcCsym), KBC D (\kbcDsym), ELBM (\elbmsym). Results represented by symbols have been subsampled for clarity.}
     	\label{fig:2dturbESR13}
\end{figure}

We consider three groups of numerical experiments. The first simulations are carried out at a moderate Reynolds number ${\rm Re}=13'134$ where the energy spectrum was peaked at $k_0 = 9$, the second group was run at ${\rm Re}=1.5\cdot 10^5$ with a different random initial velocity field, $k_0 = 30$, and the last group was simulated at a very high Reynolds number ${\rm Re}=1.6\cdot 10^6$ with the same initial field. Reynolds number is defined as ${\rm Re} = N \sqrt{(2E)}/\nu$, where $E$ is the mean initial kinetic energy. A rough estimate of the eddy turnover time is given by $t_e \approx Z^{-1/2}$ \cite{Millen}. All the simulations were run for $100\,\, t_e$ in order to observe both vortex formation, merging and decay. The characteristic figures for the initial conditions are summarized in table \ref{tab:2dturbsetup}.
As in all our simulations, Grad's approximation \eqref{eq:Grad} was used to initialize the populations, while here the gradients of velocity were estimated by central differences from the given initial field. Note that LBGK could not cope with the under-resolved cases $N=1024,2048$ at ${\rm Re}=1.6\cdot 10^6$ and ``crashed'' due to numerical instabilities where the other five methods run trouble-free.

\begin{figure}[!t]
    \centering
    		
    	\begin{subfigure}[b]{0.45\textwidth}
    		\centering
    		\includegraphics[width=\textwidth]{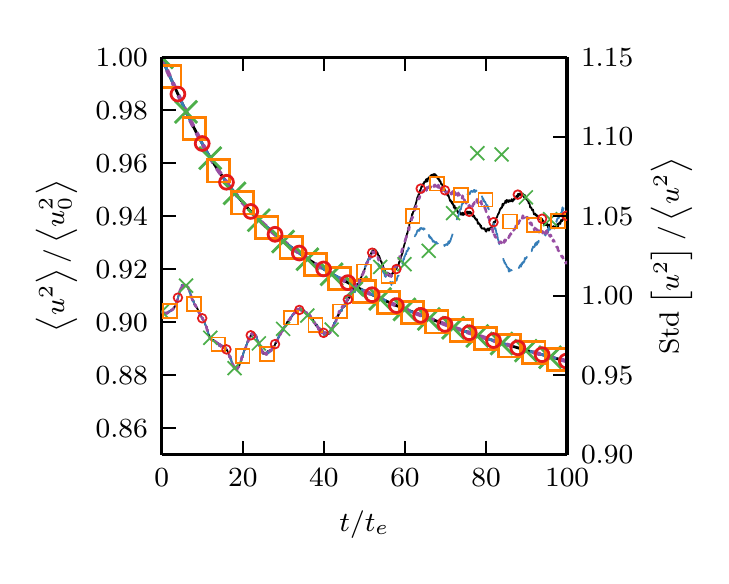}\llap{\parbox[b]{2.0in}{\textbf{(a)}\\\rule{0ex}{0.50in}}}
    	\end{subfigure}
    	\begin{subfigure}[b]{0.45\textwidth}
    		\centering
    		\includegraphics[width=\textwidth]{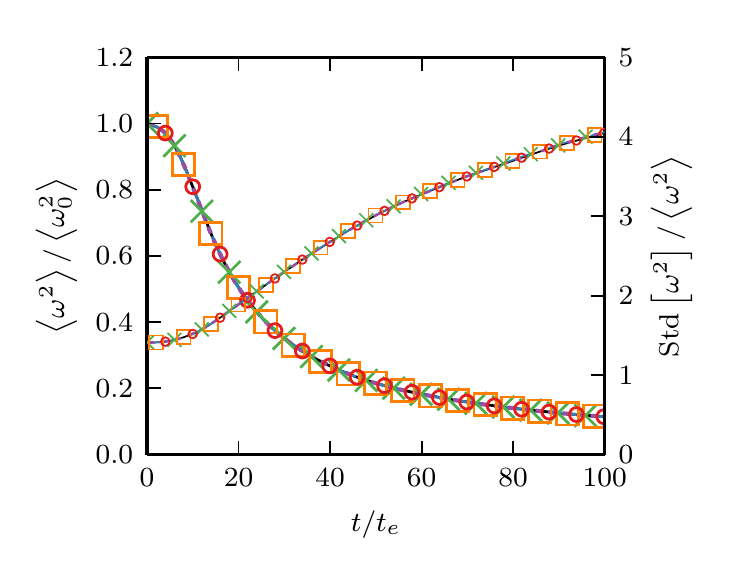}\llap{\parbox[b]{2.05in}{\textbf{(b)}\\\rule{0ex}{0.50in}}}
    	\end{subfigure}	
    	\vspace*{-0.5cm}
    	
    	\begin{subfigure}[b]{0.45\textwidth}
    		\centering
   		\includegraphics[width=\textwidth]{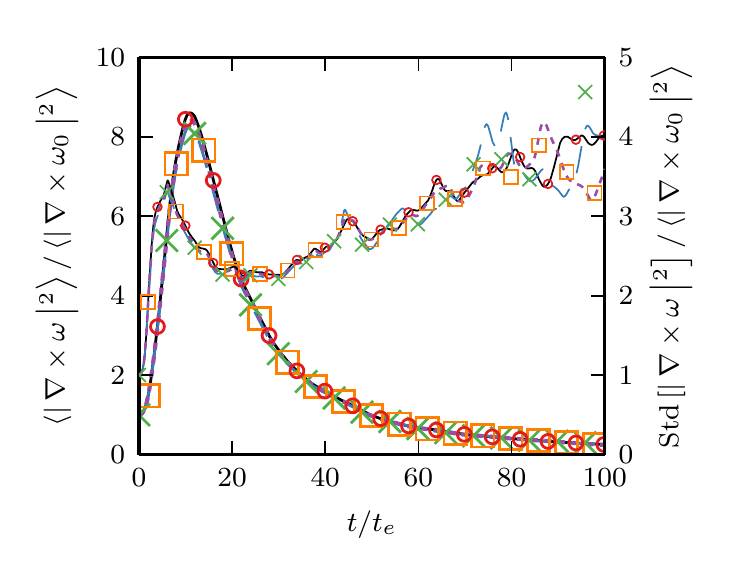}\llap{\parbox[b]{2.0in}{\textbf{(c)}\\\rule{0ex}{0.50in}}}
    	\end{subfigure}
    	\begin{subfigure}[b]{0.45\textwidth}
    		\centering
    		\includegraphics[width=\textwidth]{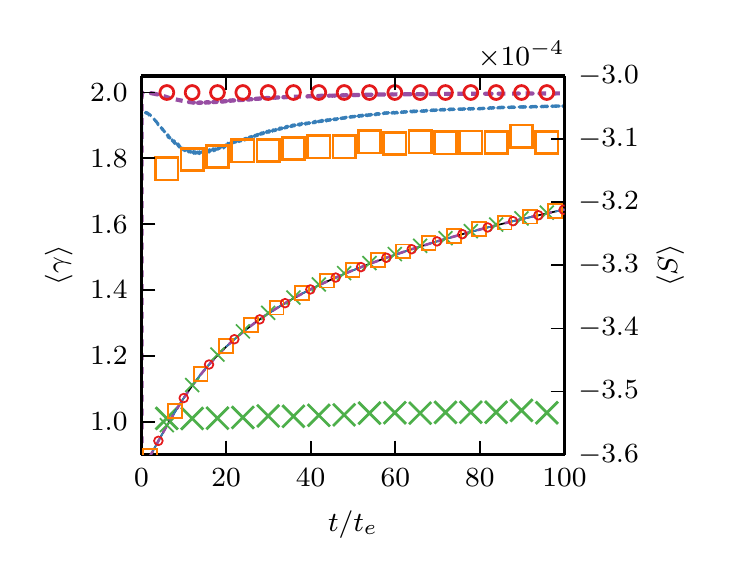}\llap{\parbox[b]{2.05in}{\textbf{(d)}\\\rule{0ex}{0.50in}}}
    	\end{subfigure}
    	\vspace*{-0.3cm}
    	
    	\caption{Evolution of kinetic energy (a), enstrophy (b), palinstrophy (c) and stabilizer $\gamma$ (d) for $Re = 1.5\cdot 10^5$ at resolution $N=4096$. Mean: left y-axis and large symbols, standard deviation: right y-axis and smaller symbols. LBGK (\lbgksym), KBC A (\kbcAsym), KBC B (\kbcBsym), KBC C (\kbcCsym), KBC D (\kbcDsym), ELBM (\elbmsym). Results represented by symbols have been subsampled for clarity.}
    	\label{fig:2dturbEnergyR150}
\end{figure}

Let us first consider the low Reynolds number case. Figure \ref{fig:2dturbvortR13} shows a comparison of the vorticity field for LBGK and KBC B and $N=1024$ at different times. The first column shows the vorticity structures at the point of maximum turbulence activity indicated by the palinstrophy evolution (see fig. \ref{fig:2dturbPalinGammaR13}). The vortices have been formed and coherent structures appear in the next shown time instance which interact with each other. The number of vortices is clearly decaying when comparing the last column of fig. \ref{fig:2dturbvortR13} with the earlier time instances. It is striking that all the models show almost the same dynamics (for brevity fig. \ref{fig:2dturbvortR13} shows only LBGK and KBC B). Except for the last time point the plots are visually hardly discriminable. Even after a long time, $t/t_e = 100$, the vorticity structures are still comparable.

The decay of enstrophy was measured and figs. \ref{fig:2dturbenergyR13} b) and d) show the expected exponential decay for resolutions $N=256$ and $N=1024$. It is apparent that the evolution of mean enstrophy is the same for all models and is almost identical among the two resolutions. The fluctuations are largely the same, although one can observe a slight flattening at later times for the KBC models compared to ELBM and LBGK for $N=256$. The evolution of energy, figs. \ref{fig:2dturbenergyR13} a) and c), shows a similar tendency; the models coincide in the mean but differ in the fluctuations for the lower resolution. As all the models are close to each other at $N=1024$ and the means of enstrophy and energy are not changing compared to $N=512$ we consider this highest resolution as resolved.

Evidence for this classification is also gathered from figure \ref{fig:2dturbPalinGammaR13} a) and c), where the mean palinstrophy and its fluctuations match for all models at the highest resolution. As stated earlier, at $t/t_e\approx 10$ we observe a peak in palinstrophy which indicates a state of high turbulence intensity. Note that the maximum value is slightly better captured by ELBM and LBGK in the low resolution case.

\begin{figure}[t]
    \centering

    	\begin{subfigure}[b]{0.45\textwidth}
    		\centering
   		\includegraphics[width=\textwidth]{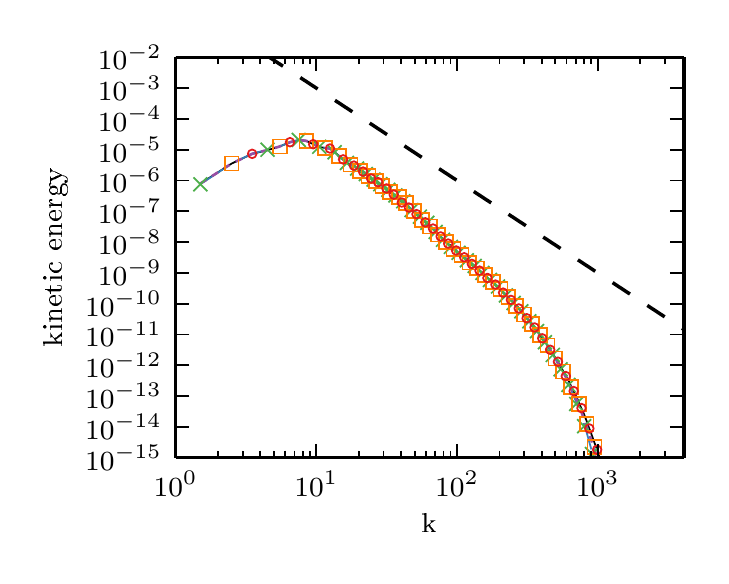}\llap{\parbox[b]{1.9in}{\textbf{(a)}\\\rule{0ex}{0.5in}}}
    	\end{subfigure}
    	\begin{subfigure}[b]{0.45\textwidth}
    		\centering
    		\includegraphics[width=\textwidth]{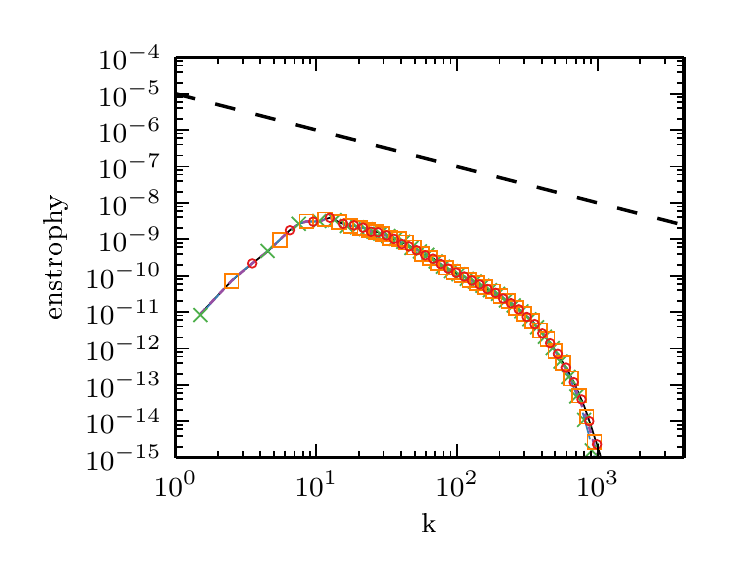}\llap{\parbox[b]{1.9in}{\textbf{(b)}\\\rule{0ex}{0.5in}}}
    	\end{subfigure}
    	\vspace*{-0.3cm}
    	
     	\caption{Energy (a) and enstrophy spectra (b) for two-dimensional turbulence at ${\rm Re} = 1.5 \cdot 10^5$ for $N=4096$ at time $t/t_e=50$. LBGK (\lbgksym), KBC A (\kbcAsym), KBC B (\kbcBsym), KBC C (\kbcCsym), KBC D (\kbcDsym), ELBM (\elbmsym). Results represented by symbols have been subsampled for clarity.}
     	\label{fig:2dturbESR150}
\end{figure}

For all the measured low-order statistical moments we have seen almost identical values, at least in the resolved case, and largely identical mean statistics overall. It is interesting to see, however, that the KBC models differ quite significantly among each other with respect to the evolution of the stabilizer $\gamma$, see fig. \ref{fig:2dturbPalinGammaR13} b) an d). Especially, KBC B is fundamentally different from the other KBC models by $\gamma$ staying close to $1$ all the time. This is consistent with the ``quasi-orthogonality'' of the $\Delta h$ and $\Delta s$ decomposition of this model demonstrated above 
(see eqs.~(\ref{eq:dsdh0B}) and (\ref{eq:dsdh1B})).
Nevertheless, the overall production of entropy is nearly identical for all the considered models.

The scaling of spectral energy and enstrophy density is shown in fig. \ref{fig:2dturbESR13} for $N=1024$ and $t/t_e=50$. Due to the moderate Reynolds number the slopes are not expected to match with the theoretical prediction for high Reynolds numbers,
however, we are not able to distinguish between the models. Thus we can conclude that for the moderate Reynolds number all methods give almost identical results and show the same dynamics.

The next simulation was carried out with a higher Reynolds number,  ${\rm Re}=1.5\cdot 10^5$, but on a larger grid. The number of initial vortices is much higher due to a different random initial condition. Here we report the results for the highest resolution, $N=4096$. For the lower resolutions, the results are similar, albeit with slightly more variance among the models with respect to the fluctuations of energy.

According to figs. \ref{fig:2dturbEnergyR150} and \ref{fig:2dturbESR150}, where evolution of kinetic energy, enstrophy and palinstrophy as well as spectral density of energy and enstrophy are reported, we observe that the six models behave almost identically. Note that for this Reynolds number LBGK did not encounter any numerical instabilities. Due to the similarities among the models and the fact that the results for $N=2048$ are not significantly different from the largest resolution, we conclude that for $N=4096$ the flow is essentially resolved. Note the differences among the models in figure \ref{fig:2dturbEnergyR150} d) for the stabilizer $\gamma$ which is in accordance with the results from the lower Reynolds number.
Although, the Reynolds number here is one order of magnitude higher than before, the scaling of energy and enstrophy is still too steep compared to the theoretical slope.

\begin{figure}[t]
    \centering
    \begin{subfigure}[b]{0.45\textwidth}
    		\centering
    		\includegraphics[width=\textwidth]{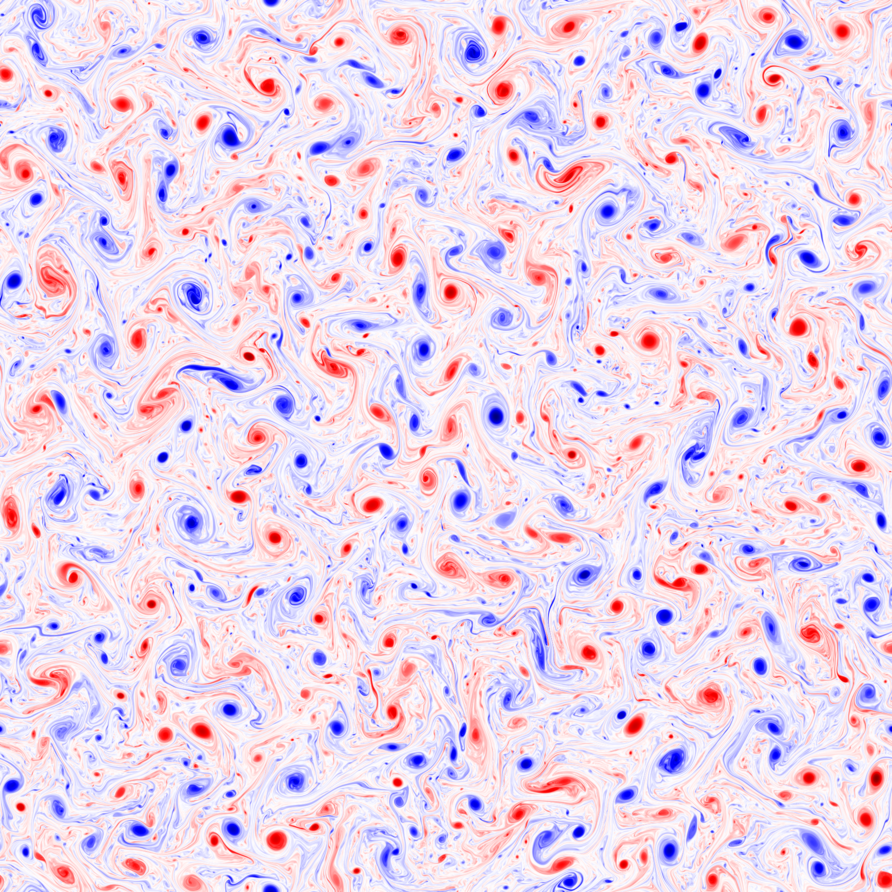}\llap{\parbox[b]{2.5in}{\textbf{(a)}\\\rule{0ex}{0.2in}}}
    	\end{subfigure}~
    	\begin{subfigure}[b]{0.45\textwidth}
    		\centering
    		\includegraphics[width=\textwidth]{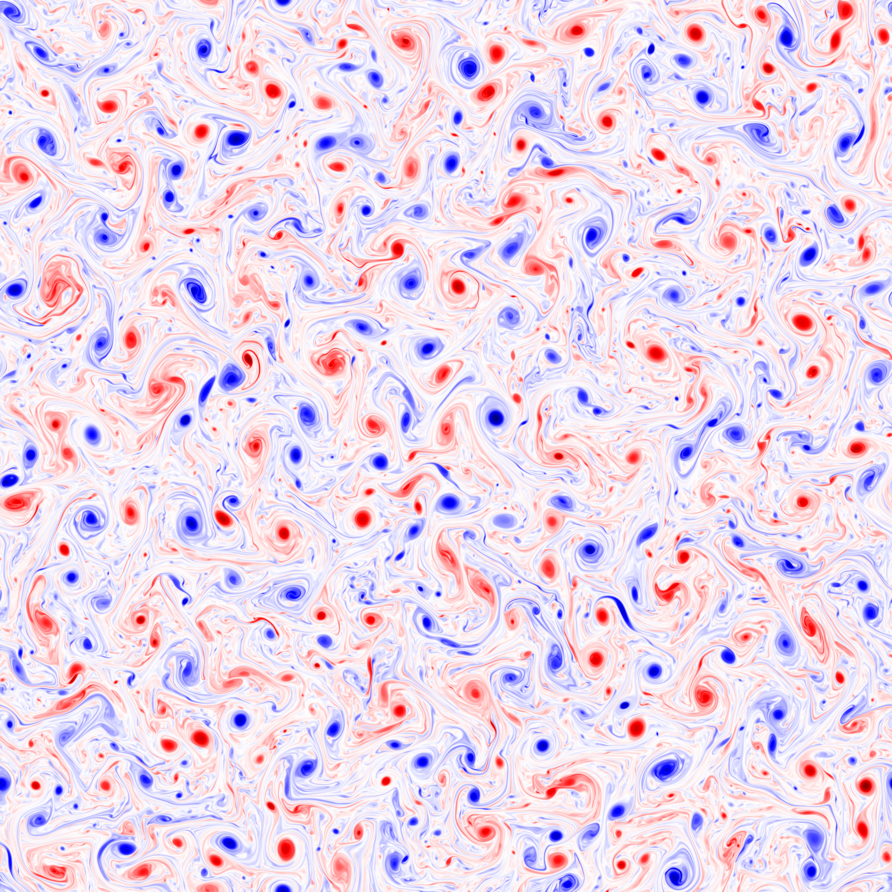}\llap{\parbox[b]{2.5in}{\textbf{(b)}\\\rule{0ex}{0.2in}}}
    	\end{subfigure}
    	
	\caption{Vorticity field for decaying two-dimensional turbulence, ${\rm Re}=1.6\cdot 10^6$, for different ELBM (a) and KBC B (b) for time $t/t_e=50$.}
    	\label{fig:2dturbvortR1600}
\end{figure}

In order to verify that the discussed models can achieve the proper scaling laws for very high Reynolds numbers, we conducted a simulation at ${\rm Re}=1.6\cdot 10^6$.
Let us first remark that for this highly turbulent regime, LBGK was not able to run with $N=1024$ and $N=2048$. There is  evidence that all of the considered grids do not fully resolve the flow as the mean statistical quantities are still slightly different among the two highest resolutions, $N=2048, 4096$, so that the six models are affected differently by the lack of resolution.
It is thus interesting to see whether the dissipation is affected by the KBC models in in the presence of under-resolution.

For this matter let us compare the ELBM model and KBC B for the time $t/t_e=50$ and $N=4096$. Figure \ref{fig:2dturbvortR1600} shows the vorticity field, accordingly. Note that the all the models produce approximately the same vortex structures up to $t/t_e \sim 20$.  Although, one can still see similarities of the structures at $t/t_e=50$, the two models produce distinctly different pictures. The number of vortices, roughly estimated by the number of vorticity patches exceeding two times the standard deviation of vorticity, is clearly different: ELBM accounts for $3440$ whereas KBC B shows $1587$ vortices. While visually the number of larger vortices seems comparable, this difference must stem from the very small structures. 

\begin{figure}[t]
    \centering
    \begin{subfigure}[b]{0.32\textwidth}
    		\centering
   		\includegraphics[width=\textwidth]{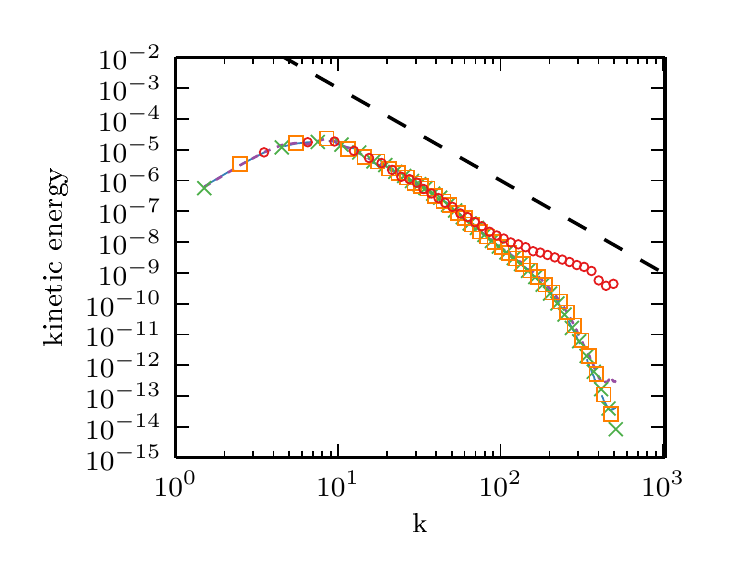}\llap{\parbox[b]{1.35in}{\textbf{(a)}\\\rule{0ex}{0.4in}}}
    	\end{subfigure}
    	\begin{subfigure}[b]{0.32\textwidth}
    		\centering
    		\includegraphics[width=\textwidth]{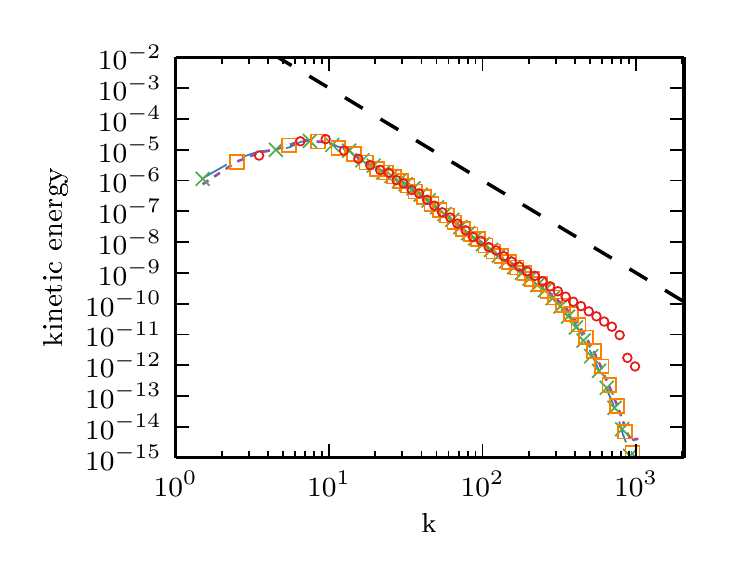}\llap{\parbox[b]{1.35in}{\textbf{(b)}\\\rule{0ex}{0.4in}}}
    	\end{subfigure}  		
    	\begin{subfigure}[b]{0.32\textwidth}
    		\centering
   		\includegraphics[width=\textwidth]{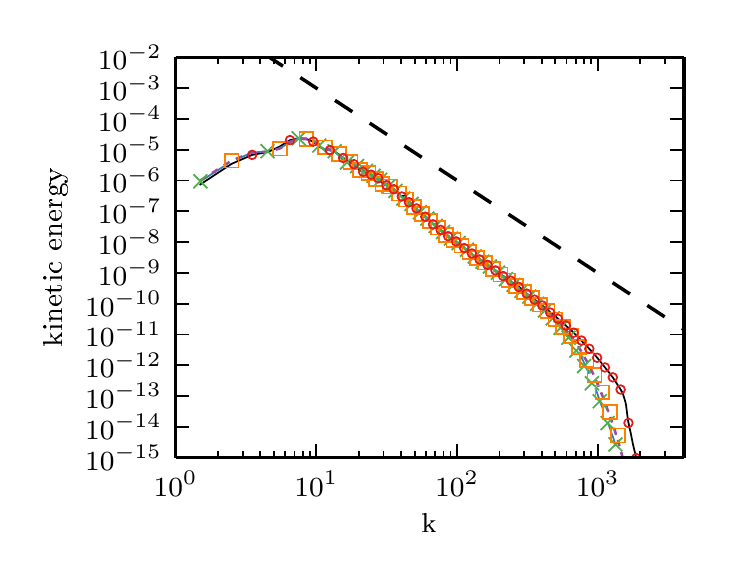}\llap{\parbox[b]{1.35in}{\textbf{(c)}\\\rule{0ex}{0.4in}}}
    	\end{subfigure}
	\vspace*{-0.5cm}
    		
    	\begin{subfigure}[b]{0.32\textwidth}
    		\centering
    		\includegraphics[width=\textwidth]{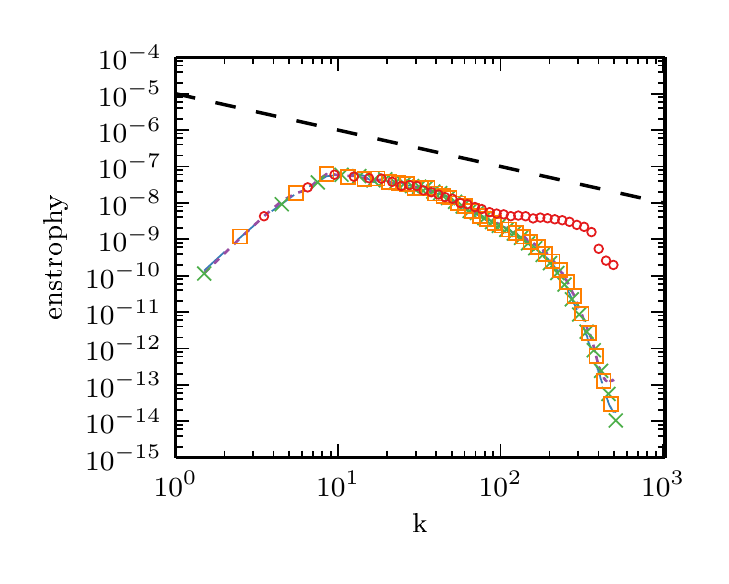}\llap{\parbox[b]{1.35in}{\textbf{(d)}\\\rule{0ex}{0.4in}}}
    	\end{subfigure}
    	\begin{subfigure}[b]{0.32\textwidth}
    		\centering
   		\includegraphics[width=\textwidth]{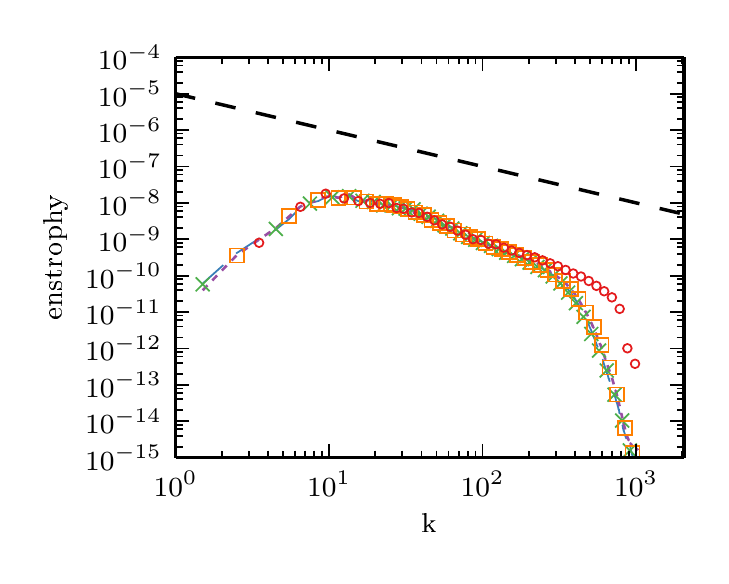}\llap{\parbox[b]{1.35in}{\textbf{(e)}\\\rule{0ex}{0.4in}}}
    	\end{subfigure}
    	\begin{subfigure}[b]{0.32\textwidth}
    		\centering
    		\includegraphics[width=\textwidth]{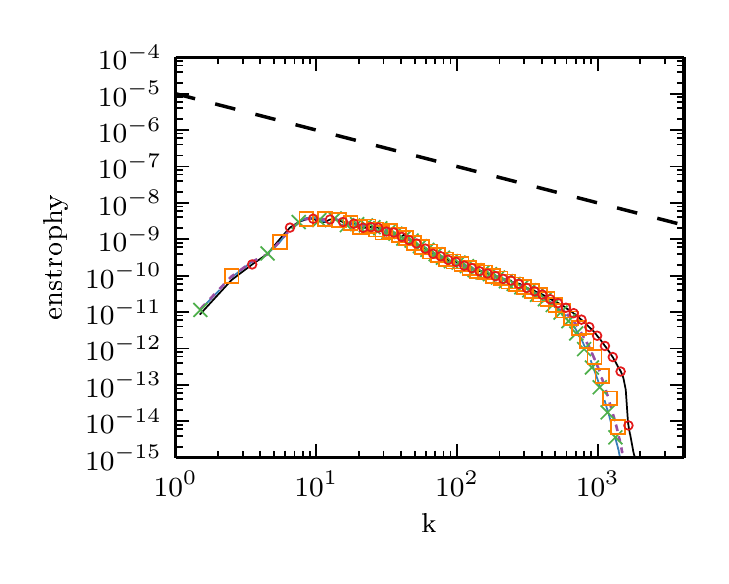}\llap{\parbox[b]{1.35in}{\textbf{(f)}\\\rule{0ex}{0.4in}}}
    	\end{subfigure}
    	\vspace*{-0.3cm}
    	
    \caption{Energy (a) - (c) and enstrophy spectra (d) - (f),for two-dimensional turbulence at ${\rm Re} = 1.6 \cdot 10^6$ for $N=1024,2048,4096$ (top to bottom) at times $t/t_e=50$. LBGK (\lbgksym), KBC A (\kbcAsym), KBC B (\kbcBsym), KBC C (\kbcCsym), KBC D (\kbcDsym), ELBM (\elbmsym). Results represented by symbols have been subsampled for clarity.}
	\label{fig:2dturbESR1600}
\end{figure}

These findings are also consistent with the energy and enstrophy spectra depicted in fig \ref{fig:2dturbESR1600}. 
While ELBM keeps more energy and enstrophy in the large wave numbers, KBC models smoothly fall off. At lower resolution, ELBM shows a bump near the largest wave numbers (see fig. \ref{fig:2dturbESR1600}) which was observed in other simulations as well. This is conjectured to be the effect of the built-in subgid model established through the fluctuating effective viscosity. It can be observed, however, that all the models capture the theoretical slope in a range of wave numbers. 

\begin{figure}[!]
    \centering
	\begin{subfigure}[b]{0.45\textwidth}
    		\centering
		\includegraphics[width=\textwidth]{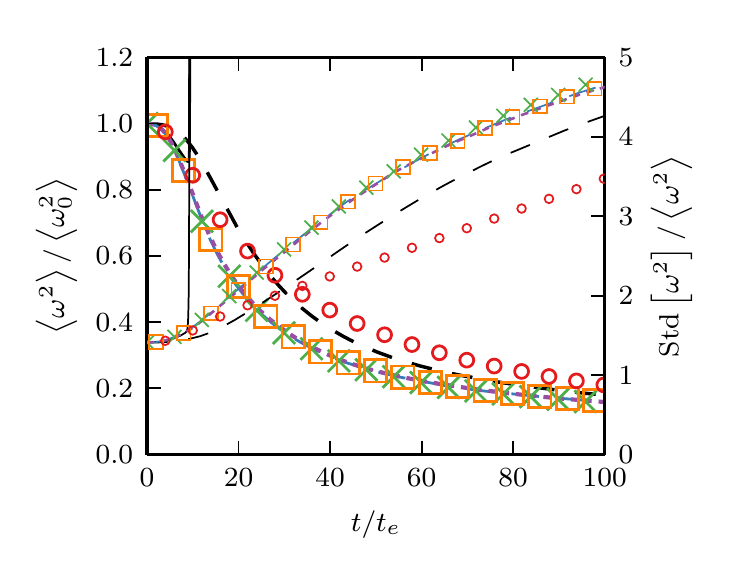}\llap{\parbox[b]{2.0in}{\textbf{(a)}\\\rule{0ex}{0.5in}}}    
    \end{subfigure}
    \begin{subfigure}[b]{0.45\textwidth}
    		\centering
    		\includegraphics[width=\textwidth]{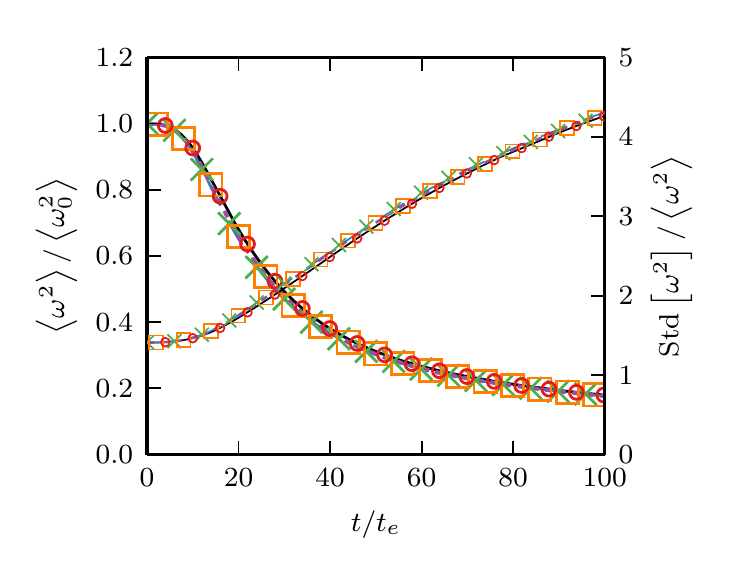}\llap{\parbox[b]{2.0in}{\textbf{(b)}\\\rule{0ex}{0.5in}}}    
    	\end{subfigure}
    	\caption{Evolution of enstrophy for $Re = 1.6\cdot 10^6$ at resolution $N=1024$ (a) and $N=4096$ (b). Mean: left y-axis and large symbols, standard deviation: right y-axis and smaller symbols. LBGK (\lbgksym), KBC A (\kbcAsym), KBC B (\kbcBsym), KBC C (\kbcCsym), KBC D (\kbcDsym), ELBM (\elbmsym). Results represented by symbols have been subsampled for clarity. In (a) the dark dashed line represents the corresponding values at $N=4096$ for LBGK.}
    	\label{fig:2dturbEnstrophyR1600}
\end{figure}

On the other hand, the enstrophy evolution, depicted in fig. \ref{fig:2dturbEnstrophyR1600} b), shows almost identical average dissipation for all the models at $N=4096$. This indicates that despite the difference in the spectra KBC models do not introduce a significantly higher dissipation, however, it seems that the flux of energy to larger scales is more dominant than in the case of ELBM. Fig. \ref{fig:2dturbEnstrophyR1600} a) shows the enstrophy decay for the resolution $N=1024$. Here, the ELBM method clearly shows a slower decay than KBC. When considering the corresponding curve for $N=4096$ and LBGK as a reference (dark dashed line) it is apparent that none of the models capture the expected rate, however, ELBM slightly under predicts the decay at later times while the KBC models show in general more dissipation (see also fig. \ref{fig:2dturbESR1600}) d) and e)).
In contrast to the smallest scales, the large and moderately small scales seem to be predicted well by all methods.

\subsection*{Stability and Performance}

In all the simulations reported in this paper as well as in preliminary tests and the simulations published in \cite{KBC} we did not encounter a single case of numerical instability using the KBC models, even when running very under-resolved simulations for high Reynolds numbers. 
The ELBM method is non-linearly stable and thus it is not surprising that it runs trouble-free for all the set-ups. LBGK on the other hand failed numerically in presence of under-resolution.

The KBC models introduce additional computational overhead in order to compute additional moments and the estimate for the stabilizer $\gamma$ which accounts for not more than a factor of $2 - 2.5$ for both two and three dimensions.

\section{Conclusions}

In this work we studied four variations of entropy based multi relaxation models of the recently introduced KBC family. We reviewed the details of the entropic stabilization and described the four models in detail. The recovery of the Navier-Stokes equations was demonstrated to second order for all the KBC models.
A detailed 
comparison with LBGK and ELBM was carried out at various grid resolutions and Reynolds numbers for different two-dimensional flows. Second order rate of convergence is numerically confirmed for all the models studied herein.

It must be stressed that the entropic models, KBC and ELBM, were stable (in contrast to LBGK) for all the considered cases here;
despite of under-resolution and high Reynolds numbers (e.g. ${\rm Re}= 1.6\cdot 10^6$ for a grid of $1024\times 1024$). 

The accuracy of various models was discussed, in particular, for under-resolved simulations.
We first remark, that all the models converge to the LBGK solution for resolved flows and that, in general, the difference among the models under consideration are small. At moderately high Reynolds number, the KBC models perform better than LBGK, which suffers from numerical instabilities, and ELBM which produces spurious vortices in the case of the double shear layer.

The simulation of two-dimensional turbulence was chosen as a benchmark to assess various statistics for high Reynolds numbers. 
While covering a range of different resolutions and Reynolds numbers, the KBC models were shown to capture the expected scaling laws for energy and enstrophy spectra as well as ELBM (and LBGK for sufficiently large resolutions). There is indication that the KBC models produce less small structures than ELBM (and LBGK) for the under-resolved cases, however, the decay of enstrophy is only slightly accelerated in the under-resolved case for high Reynolds numbers.  On the other hand ELBM tends to somewhat amplify the appearance of small structures which lead to a slight over-representation of enstrophy (and energy) content at large wave numbers but only for very coarse resolutions.
In summary, 
all the KBC models considered here have the correct limit of LBGK for resolved simulations, but more importantly, they capture all low order statistics such as averages and fluctuations of kinetic energy, enstrophy and palinstrophy as well as the spectral densities for energy and enstrophy extremely well despite severe under-resolution. 

Among the KBC variations, model B was demonstrated to be more accurate in the under-resolved shear layer case. The significant difference with respect to the evolution of the stabilizer $\gamma$ for model B can be explained by the orthogonality of the entropic scalar product $\langle\Delta s{|}\Delta h\rangle$ in the leading and first order terms in velocity powers.

In general, we show that by keeping the kinematic (shear) viscosity coefficient constant (in contrast to ELBM) the presented method is extremely stable and produces accurate results in presence of under-resolution (similar to ELBM). Minor differences in performance among the different KBC versions are observed for different simulations, however, all KBC models are much more stable than LBGK.

It has been demonstrated in \cite{KBC} that also for three dimensional flows in presence of complex walls low-order statistics can be captured well using KBC. In a further publication we will address the issue of boundary conditions for KBC models in both two and three dimensions.

\section*{Acknowledgement}

This work was supported by the European Research Council (ERC) Advanced Grant No. 291094-ELBM. 
Computational resources at the Swiss National Super Computing Center CSCS were provided under the grant S492.

\bibliographystyle{plain}
\bibliography{kbc_2d_standalone}

\end{document}